\documentclass[10pt,twocolumn,longbibliography]{revtex4-1}

\usepackage{amsmath, amsfonts, amssymb}     %
\usepackage{graphicx}                       %
\usepackage{xfrac}                          %
\usepackage{grffile}                        %
\usepackage{color,colortbl}

\usepackage[colorlinks, citecolor=blue, urlcolor=red, linkcolor=blue, hypertexnames=false]{hyperref}
\def\arraystretch{1.5}%

\makeatletter

\def\CT@@do@color{%
  \global\let\CT@do@color\relax
  \@tempdima\wd\z@
  \advance\@tempdima\@tempdimb
  \advance\@tempdima\@tempdimc
  \advance\@tempdimb\tabcolsep
  \advance\@tempdimc\tabcolsep
  \advance\@tempdima2\tabcolsep
  \kern-\@tempdimb
  \leaders\vrule
  \hskip\@tempdima\@plus  1fill
  \kern-\@tempdimc
\hskip-\wd\z@ \@plus -1fill }
\makeatother

\renewcommand{\eqref}{Eq.~\ref}

\newcommand{\tsr}[1] {\mathbf{#1}}
\newcommand{\vtr}[1] {\mathbf{#1}}

\newcommand{\gdot}[0] {\dot{\gamma}}
\newcommand{\gdotb}[0] {\bar{\dot{\gamma}}}
\newcommand{\gdotbar}{\overline{\dot{\gamma}}}
\newcommand{\degree}{\ensuremath{^\circ}}

\newcommand{\etas}{\eta_{\rm s}}
\newcommand{\etaa}{\eta_{\rm a}}

\newcommand{\dob}[0] {\Delta_{\hspace{-1pt}\gdot}}
\newcommand{\dobc}[0] {\Delta^{\hspace{-1pt}c}_{\hspace{-1pt}\gdot}}
\newcommand{\dos}[0] {\Delta^{\sigma}}

\newcommand{\lm}[0]{\ensuremath{\ell_\mu}}

\newcommand{\eg}{e.g.,~}

\newcommand{\etal}{\textit{et al.}~}

\newcommand{\be}{\begin{equation}}
\newcommand{\ee}{\end{equation}}
\newcommand{\bea}{\begin{eqnarray}}
\newcommand{\eea}{\end{eqnarray}}

\newcommand{\xhat}{\hat{\mathbf{x}}}
\newcommand{\yhat}{\hat{\mathbf{y}}}
\newcommand{\zhat}{\hat{\mathbf{z}}}

\newcommand{\vecv}[1]{\mathbf{{#1}}}
\newcommand{\tens}[1]{\mathbf{{#1}}}

\begin{document}
\title{Interplay of edge fracture and shear banding in complex fluids}
\author{Ewan J. Hemingway and Suzanne M. Fielding}
\affiliation{Department of Physics, Durham University, Science Laboratories,
  South Road, Durham DH1 3LE, UK}
\date{\today}
\begin{abstract}

We explore theoretically the interplay between shear banding and edge
fracture in complex fluids, by performing a detailed simulation study
within two constitutive models: the Johnson-Segalman model and the
Giesekus model. We consider separately parameter regimes in which the
underlying constitutive curve is monotonic and non-monotonic, such
that the bulk flow (in the absence of any edge effects) is
respectively homogeneous and shear banded.  Phase diagrams of the
levels of edge disturbance and of bulk (or quasi-bulk) shear banding
are mapped as a function of the surface tension of the fluid-air
interface, the wetting angle where this interface meets the walls of
the flow cell, and the imposed shear rate. In particular, we explore
in more detail the basic result recently announced in
Ref.~\cite{hemingway2018edge}: that precursors to edge fracture can
induce quasi-bulk shear banding. We also appraise analytical
predictions that shear banding can induce edge
fracture~\cite{Skorski2011}.  Although a study of remarkable early
insight, Ref.~\cite{Skorski2011} made some strong assumptions about
the nature of the ``base state'', which we examine using direct
numerical simulation. The basic prediction that shear banding can
cause edge fracture remains valid, but with qualitatively modified
phase boundaries.
\end{abstract}

\maketitle

\section{Introduction}

In many complex fluids, a state of initially homogeneous bulk shear
flow is unstable to the formation of coexisting bands of differing
shear rate, with layer normals in the flow-gradient direction. This
phenomenon, which is called shear
banding~\cite{recentReview,ISI:000254405700003,ISI:000254405700004,ISI:000344142000001},
has been observed in wormlike micellar
surfactants~\cite{britton1997c}, lyotropic lamellar
phases~\cite{Salmonetal2003a}, triblock
copolymers~\cite{Mannevilleetal2007a}, telechelic
polymers~\cite{michel2001unstable}, star polymers~\cite{rogers2008},
clays~\cite{martin-sm-8-6940-2012,Coussotetal2002c},
emulsions~\cite{Coussotetal2002c} and (subject to
controversy~\cite{Li2013,ISI:000339141600010})
monodisperse linear entangled
polymers~\cite{ISI:000286154200002,ISI:000254645200053}. It is thought
to stem from a non-monotonicity in the underlying bulk constitutive
relation between shear stress and shear rate for homogeneous shear,
$\sigma(\gdot)$. A state of initially homogeneous shear is linearly
unstable to the formation of shear bands in the regime of negative
slope, $d\sigma/d\gdot<0$~\cite{Yerushalmi1970}. The steady state flow
curve of shear stress as a function of gap-averaged shear rate,
$\sigma(\gdotbar)$, then displays a plateau over the range of shear
rates for which the flow is banded.

\begin{figure}[!b]
  \includegraphics[width=\columnwidth]{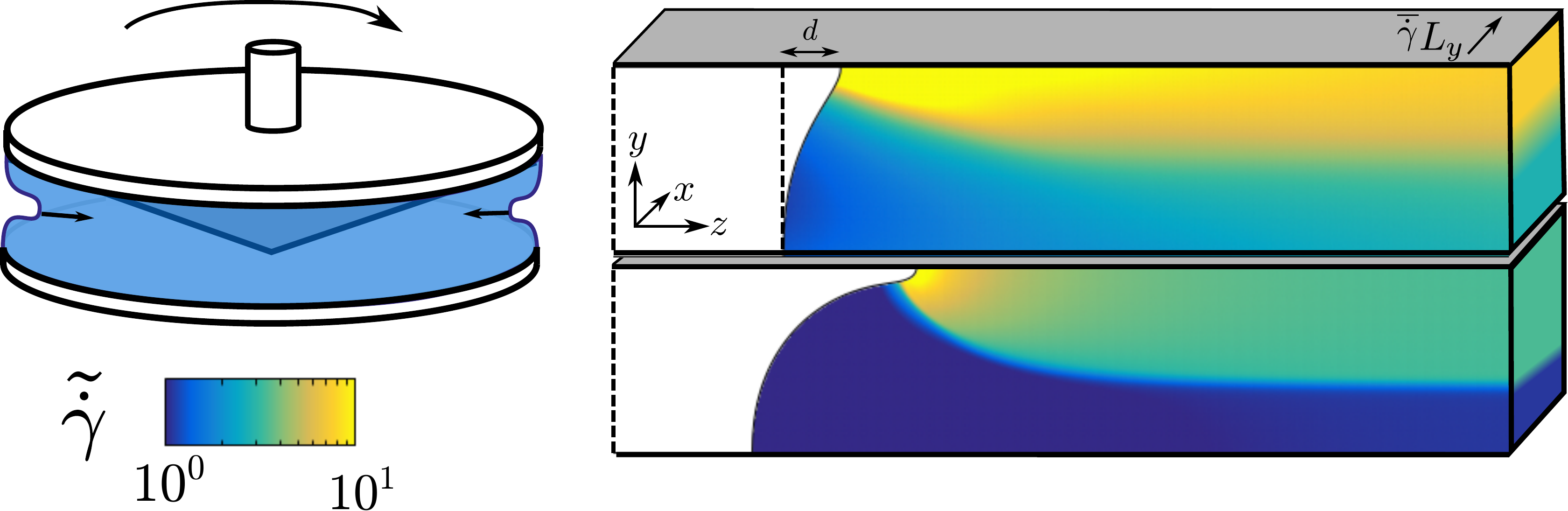}
  \caption{{\bf Top left:} schematic of a cone-plate device. {\bf Top right:} snapshot of a state in which an edge disturbance induces apparent shear banding that invades far into the
  bulk. (Giesekus model with a monotonic constitutive curve, $\gdotb =
  4.7, \etas = 0.006, \Gamma = 0.24$.) {\bf Bottom right:} snapshot of a state in which shear-banding induces a
  significant edge disturbance. (Johnson-Segalman model with a
  non-monotonic constitutive curve, $\gdotb = 2.5, \etas = 0.1, \Gamma
  = 0.11$.)  {\bf Bottom left:} colourscale
  of invariant shear rate in snapshots.
} 
  \label{fig:schematic}
\end{figure}

Many flow instabilities depend not only on bulk rheology, however, but
also on the boundary conditions where the fluid meets the walls of the
flow cell and/or the outside air.  In the cone-plate device sketched
in Fig.~\ref{fig:schematic} (top left), for example, the fluid (shown
in blue) has an interface with the air (white). When a highly
viscoelastic fluid is strongly sheared, this free surface can
destabilise to give a more complicated edge profile, with an
indentation that invades the fluid bulk.  Some portion of the sample
can even be ejected from the device.  This phenomenon, which is called
`edge-fracture', renders accurate rheological measurements very
difficult. Anecdotal reports of edge fracture pervade the experimental
literature. Detailed studies can be found in
Refs.~\cite{lee1992does,inn2005effect,sui2007instability,schweizer2008departure,mattes2008analysis,jensen2008measurements,dai2013viscometric}.

Pioneering theoretical work by Tanner
\etal~\cite{tanner1983shear,keentok1999edge} identified the second normal stress
difference in the sheared fluid, $N_2(\gdot)$, as a key factor in
driving edge fracture, and proposed the criterion for the onset of
edge fracture to be $|N_2|>2\Gamma/3R$, where $\Gamma$ is the surface
tension of the fluid-air interface, and $R$ is the radius of a
pre-assumed indentation in the interfacial profile. Recently,
Hemingway et al.~\cite{Hemingway2017,hemingway2019edge} derived an
updated criterion:
\be
\frac{\sigma}{2}\frac{d|N_2(\gdot)|}{d\gdot}\bigg/\frac{d\sigma}{d\gdot}>\frac{2\pi\Gamma}{L_y},
\label{eqn:criterion}
\ee
and showed it to be in full agreement with numerical
simulations. (Here $L_y$ is the size of the gap between the rheometer
plates.)  This criterion marks the transition with increasing flow
rate (which determines the LHS of the above inequality) or decreasing
surface tension (on the RHS) from a state in which the fluid-air
interface is undisturbed by flow to one in which it can become
significantly deformed, but remains in tact overall: {\it i.e.,} a
partially edge fractured state.  Full edge fracture, explored by
direct numerical simulation~\cite{Hemingway2017,hemingway2019edge},
follows for even stronger flows or smaller surface tensions, with the
fluid then completely dewetting the walls.

So far, we have discussed shear banding as a purely bulk instability,
and edge fracture as an interfacial instability, implicitly suggesting
that the two instabilities act independently of each other, and
without any interplay between them. Recent
work~\cite{hemingway2018edge}, however, showed that even only
modest deformations of the fluid-air interface can lead to strong
secondary flows in the fluid. For a material with a relatively flat
bulk constitutive curve $\sigma(\gdot)$, these can take the form of
apparent shear bands that invade deep into the bulk, even for a curve
that is monotone increasing, $d\sigma/d\gdot > 0$. In this way,
precursors of the {\em interfacial} instability of edge fracture can
precipitate quasi-{\em bulk} shear banding.  A simulation snapshot of
this phenomenon is shown in Fig.~\ref{fig:schematic}, top right. To
paraphrase this scenario: ``(precursors to) edge fracture can induce
(apparent) shear banding''.

Conversely, in many fluids that show true bulk shear banding (even in
the absence of any edge effects), the fluid-air interface often
destabilises above a critical shear rate on the stress plateau of the
flow curve, leading to the sample being ejected from the flow
device~\cite{Britton1999,Berret1997}. This phenomenon was studied
theoretically by Skorski and Olmsted~\cite{Skorski2011}, who showed
that the fluid-air interface of a shear banded sample must always be
at least partially disturbed, with meniscus curvatures set by the jump
in second normal stress across the interface between the bulk bands. A
simulation snapshot of this phenomenon is shown in
Fig.~\ref{fig:schematic}, bottom right. As the relative fraction of the
rheometer gap taken up by each band changes with the overall imposed
shear rate, one of the bands can develop a width that cannot support
the meniscus curvatures demanded by this second normal stress balance,
leading to full edge fracture.  To paraphrase this scenario ``shear
banding can induce edge fracture''.

For many years, an outstanding question has been whether the
underlying constitutive curve of monodisperse linear entangled
polymers is monotonic or non-monotonic, and so whether a steady
applied shear flow should be homogeneous or shear banded. Because the
underlying constitutive curve cannot be accessed experimentally in a
shear banding system, this question must be settled by explicit
velocimetry of the flow field. Tapadia and Wang ~\cite{Tapadia2006}
gave evidence for steady state banding, suggesting a non-monotonic
constitutive curve.  In contrast, Hu et al.~\cite{Hu2007} found
banding only transiently during shear startup, with homogeneous shear
recovered at longer times, suggesting a monotonic constitutive
curve. Work on more highly entangled samples did however report
long-lived bands in some runs, but not others~\cite{Hu2010}, even when
repeated for the same imposed flow rates. Edge fracture was discussed
as a possible source of this variability. Significant possible edge
fracture in experiments investigating shear banding in entangled
polymers was likewise discussed in
Refs.~\cite{Sui2007,Li2013,Wang2014,Li2014a}.

We have seen, then, that ``(precursors to) edge fracture can induce
(apparent) shear banding'', that ``shear banding can induce edge
fracture'', and that the relative roles of shear banding and edge
fracture remain unclear in some experiments, particularly on linear
entangled polymers. Theoretical work is clearly, therefore, needed to
disentangle the relative contributions of shear banding and edge
fracture in highly viscoelastic fluids, and to assess any interplay
between them.

The aim of the present manuscript is to explore theoretically this
interplay between shear banding and edge fracture, by performing a
detailed simulation study within two constitutive models: the
Johnson-Segalman model~\cite{johnson1977model} and the Giesekus
model~\cite{giesekus1982simple}. Considering separately parameter
regimes in which the underlying constitutive curve is monotonic and
non-monotonic, we shall map out phase diagrams of the levels of edge
disturbance and of bulk (or quasi-bulk) banding as a function of the
surface tension of the fluid-air interface, the wetting angle where
this interface meets the walls of the flow cell, and the imposed shear
rate. In particular, we shall explore in more detail the basic result
recently announced in Ref.~\cite{hemingway2018edge}: that
``(precursors to) edge fracture can induce (apparent) shear
banding''. We also carefully appraise the analytical predictions that
``shear banding can cause edge fracture'': although a study of
remarkable early insight, Ref.~\cite{Skorski2011} made some strong
assumptions about the nature of the ``base state'', which we will now
examine using direct numerical simulation.

The paper is structured as follows. In Sec.~\ref{sec:models} we
discuss the theoretical models to be used throughout the paper. In
Sec.~\ref{sec:geometry} we introduce the flow geometry that we
simulate, along with the boundary conditions, initial conditions, and
choice of parameter values.
We then present our results, starting in Sec.~\ref{sec:results1} with
the case of (apparent) shear banding induced by disturbances at the
fluid-air interface, for a fluid with a monotonic though relatively
flat constitutive curve. We then discuss the case of edge fracture
induced by bulk banding, given a non-monotonic constitutive curve, in
Sec.~\ref{sec:results2}. Finally Sec.~\ref{sec:conclusions} gives our
conclusions and perspectives for future work.

\section{Models}
\label{sec:models}

We work in the zero Reynolds number limit of inertialess flow, in
which the total stress in any element of fluid or air,
$\tens{T}(\vecv{r},t)$ (with $\vecv{r}$ position and $t$ time), must
obey the condition of force balance:
\be
\nabla.\tens{T}=0.
\ee
In any element inside the fluid, $\tens{T}$ comprises the sum of an
isotropic contribution characterised by a pressure $p(\vecv{r},t)$, a
Newtonian contribution with a viscosity $\etas$, and a viscoelastic
contribution $\tens{\Sigma}(\vecv{r},t)$ arising from the polymer
chains, wormlike micelles, etc. The condition of force balance then
reads:
\be
\etas\nabla^2\vecv{v}+\nabla.\tens{\Sigma}-\nabla p=0,
\ee
in which 
$\vecv{v}(\vecv{r},t)$ is the fluid velocity.

The condition of force balance in the air demands:
\be
\etaa\nabla^2\vecv{v}-\nabla p=0.
\ee

We assume the flow to be incompressible, such that the velocity field
$\vecv{v}(\vecv{r},t)$ obeys
\be
\nabla.\vecv{v}=0.
\ee
Enforcing this condition determines the pressure, $p(\vecv{r},t)$.

We consider two different constitutive models for the dynamics of the
viscoelastic stress. The first is the Johnson-Segalman
model~\cite{johnson1977model}, in which:
\begin{widetext}
\be
\partial_t\tens{\Sigma}+\vecv{v}.\nabla\tens{\Sigma} = \left(\tsr{\Sigma}\cdot \tsr{\Omega} - \tsr{\Omega}\cdot \tsr{\Sigma} \right) + a\left(\tsr{D} \cdot \tsr{\Sigma} + \tsr{\Sigma} \cdot \tsr{D}\right) + 2G(\phi)\tens{D} - \frac{1}{\tau(\phi)}\tsr{\Sigma}+\frac{\ell^2}{\tau(\phi)}\nabla^2\tens{\Sigma}.
\label{eqn:veceJS}
\ee
\end{widetext}
Here $\tens{D}=\tfrac{1}{2}(\nabla \vecv{v} + \nabla
\vecv{v}^T)$ and $\tens{\Omega}=\tfrac{1}{2}(\nabla \vecv{v} - \nabla
\vecv{v}^T)$ are respectively the symmetric and antisymmetric parts of the strain rate 
tensor, $\nabla \vecv{v}_{\alpha\beta}=\partial_\alpha v_\beta$.  The
parameter $a$, which lies in the range $-1 \le a\le 1$, describes a
slippage of the viscoelastic component relative to the solvent.  In
stationary homogeneous simple shear flow, the constitutive curve
$T_{xy}(\gdot)=\Sigma_{xy}(\gdot)+\etas\gdot$ is a non-monotonic
function of the imposed shear rate for $|a|<1$ and $\etas < 1/8$. In
this regime, a state of steady shear flow is shear banded. For $\etas
> 1/8$, in contrast, a steady state shear flow must be homogeneous, at
least in the absence of any edge disturbances. The second normal
stress, $N_2(\gdot)=\Sigma_{yy}-\Sigma_{xx}$, is negative, and scales as
$-\gdot^2$ in the limit $\gdot\to 0$.

The second constitutive model that we shall study is the Giesekus model~\cite{giesekus1982simple}, in which:
\begin{widetext}
\be
\partial_t\tens{\Sigma}+\vecv{v}.\nabla\tens{\Sigma} = \left(\tsr{\Sigma}\cdot \tsr{\Omega} - \tsr{\Omega}\cdot \tsr{\Sigma} \right) + \left(\tsr{D} \cdot\tsr{\Sigma} + \tsr{\Sigma}\cdot \tsr{D}\right) + 2G(\phi)\tens{D} - \frac{1}{\tau(\phi)}\tsr{\Sigma}-\frac{\alpha}{\tau(\phi)}\tsr{\Sigma}\cdot\tsr{\Sigma}+\frac{\ell^2}{\tau(\phi)}\nabla^2\tens{\Sigma}.
\label{eqn:veceG}
\ee
\end{widetext}
The parameter $\alpha$ in this equation captures an increase in the
rate of stress relaxation when the polymer chains are more strongly
aligned. In a stationary homogeneous shear flow, the viscoelastic
shear stress $\Sigma_{xy}(\gdot)$ is a non-monotonic function of
$\gdot$ for $\alpha > 1/2$, and monotonic for
$\alpha<1/2$~\cite{giesekus1982simple}.  The second normal stress
scales as $-\gdot^2$ at low shear rates, as in the Johnson-Segalman
model.

The spatial gradient terms prefactored by $\ell$ in
Eqns.~\ref{eqn:veceJS} and~\ref{eqn:veceG} ensure that the interface
between any shear bands has a slightly diffuse thickness that scales as
$\ell$~\cite{lu2000effects}.

We model the coexistence of fluid and air using a phase field
$\phi(\vecv{r},t)$, which obeys Cahn-Hilliard 
dynamics~\cite{bray1994advances}
\begin{align}
  \partial_t \phi + \vecv{v}.\nabla \phi &= M \nabla^2 \mu.
  \label{eq:SI_CH}
\end{align}
Here $M$ is a constant molecular mobility. The chemical potential
\begin{align}
  \mu &= G_\mu \left(-\phi + \phi^3 - \lm^2 \nabla^2 \phi\right),
  \label{eqn:freeEnergy}
\end{align}
in which the constant parameter $G_\mu$ determines the free energy of
demixing per unit volume.  This captures the coexistence of a fluid
phase, in which $\phi = 1$, with an air phase, in which $\phi =
-1$. The elastic modulus $G$ and relaxation time $\tau$ in the
constitutive equations~\ref{eqn:veceJS} and~\ref{eqn:veceG} are
functions of $\phi$, with $G(\phi=1)=1$, $\tau(\phi=1)=1$ and
$G(\phi=-1)=0$, with $\tau(\phi=-1)=0.002$, such that
viscoelastic stresses arise only in the fluid.

The fluid-air interface has a slightly diffuse thickness that scales
as the parameter $\lm$. This is needed to capture the motion of the
contact line where the fluid-air interface meets the walls of the flow
cell~\cite{kusumaatmaja2016moving}. Gradients in $\mu$ across the
interface contribute an additional source term $-\phi \nabla
\mu$ to the force balance condition, capturing the effects of the interfacial surface tension. The surface tension is
\begin{equation}
  \Gamma = \frac{2 \sqrt{2}}{3} G_\mu \lm.
  \label{eq:SI_surface_tension}
\end{equation}

\section{Flow geometry, boundary conditions and initial conditions}
\label{sec:geometry}

\begin{figure}[!b]
  \includegraphics[width=0.45\textwidth]{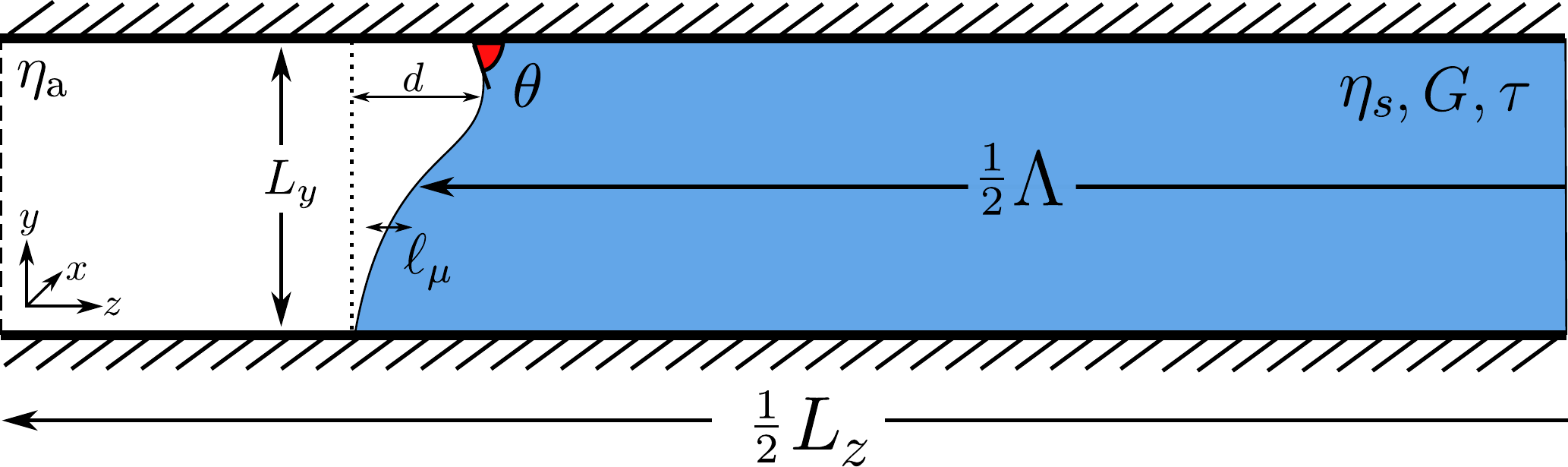}
  \caption{Flow geometry to be simulated: a planar slab of fluid
  sandwiched between hard flat parallel plates at $y=\pm L_y/2$. 
  The flow is effected by moving the top plate into the page at speed 
  $\gdot L_y$, with the bottom plate stationary. Accordingly, the base 
  state flow (before any secondary flows arise due to the instability 
  of interest here) is of the general form $\vtr{v} = (v_x(y,z),0,0)$.
   The  fluid (shown in blue) meets the air (white) at the fluid-air
  interface. Only
  the left half of the box is shown: an equivalent fluid-air interface
  in the right half is not shown.  The symbols are defined in the main
  text.}
\label{fig:geometry}
\end{figure}

We ignore any complications of slight streamline curvature and stress
heterogeneity that are present in many common experimental shear cells
(cone-plate, plate-plate, cylindrical Couette, etc.), simulating
instead a planar slab of fluid between flat parallel plates. The fluid
is sheared at an overall imposed rate $\gdotbar$ by moving the top
plate at speed $\gdot L_y$ along $\xhat$. See Fig.~\ref{fig:geometry},
in which the flow-gradient direction, $\yhat$, and vorticity direction
$\zhat$, are respectively vertical and horizontal, with the flow
direction, $\xhat$, into the page. We assume translational invariance
along $\xhat$, setting $\partial_x$ of all quantities equal to zero,
performing two-dimensional (2D) simulations in the $y-z$ plane
sketched, with all snapshots below also shown in this plane. The
velocity vector and stress tensor are nonetheless fully 3D objects.

The simulation box has length $L_z$ in the vorticity direction, with
periodic boundary conditions in that direction.  At the walls of the
flow cell in the gradient direction $y$, we assume conditions of no
slip and no permeation for the fluid velocity $\vecv{v}$. For the
viscoelastic stress $\tens{\Sigma}(\vecv{r},t)$, we assume zero
gradient:
\be
\vecv{n}\cdot\nabla \,\tens{\Sigma}=0,
\ee
in which $\vtr{n}$ is the unit outward wall normal. For the phase
field~\cite{yue2010sharp,Dong2012}, we assume
\begin{align}
  \vtr{n}\cdot \nabla \mu &= 0,\\
  \vtr{n}\cdot \nabla \phi &= \frac{-1}{\sqrt{2} \lm} \cos{\theta} \left(1 - \phi^2\right).
  \label{eq:SI_CH_BCs}
\end{align}
In the absence of flow, the contact angle at which the fluid-air
interface meets the walls of the flow cell is then given by the angle
$\theta$ in the second of these equations: $\theta=90^\circ$ gives a
vertical interface, $\theta>90^\circ$ gives an interface convex into
the air, and $\theta<90^\circ$ concave. In Ref.~\cite{hemingway2019edge}
we showed that while the steady state shape of the perturbed meniscus can depend 
on wetting angle, other quantitative predictions (\eg the location of phase boundaries) 
are relatively robust to changes in $\theta$.

\begin{table*}[!t]
  \def\arraystretch{1.25}
  \resizebox{\linewidth}{!}{%
    \noindent\begin{tabular}{| l | l | l | p{3.5cm} | l |}
      \hline
      \textbf{parameter} & \textbf{description} &  \textbf{dimension} & \textbf{value} & \textbf{notes}\\
      \hline
        $L_y$ & channel width & $[L]$ & 1.0 & unit of length\\
            $G$ & polymer modulus & $[G]$ & 1.0 & unit of stress  \\
      $\tau$ & polymer relaxation time & $[T]$ & 1.0 & unit of time  \\
      \hline
        $\gdot$ & applied shear-rate & $[T]^{-1}$ & $10^{-1}\to 10^2$ & important quantity to be varied\\
        $\theta$ & equilibrium contact angle & $[1]$ & $30\degree \to 150\degree$ & important quantity to be varied \\
        $\Gamma$ & surface tension & $[G][L]$ & $0.01 \to 1.0$ & important quantity to be varied\\
 $\etas$ & solvent viscosity (JS) & $[G][T]$ & 0.025 (strongly), 0.1 (marginally) banded & small viscosity ratio $\etas/G\tau$\\
      $\etas$ & solvent viscosity (Gk) & $[G][T]$ & $0.006 \to 0.02$ & small viscosity ratio $\etas/G\tau$\\
        \hline
          $a$ & slip parameter (JS) & $[1]$ & $0.3$ & sets dependencies $\Sigma_{xy}(\gdot), N_2(\gdot)$  \\
      $\alpha$ & anisotropy parameter (Gk) & $[1]$ & $0.8$  & sets dependencies $\Sigma_{xy}(\gdot), N_2(\gdot)$ \\
\hline
      $\etaa$ & air viscosity (JS) & $[G][T]$ & $0.01$ & small air viscosity $\etaa/G\tau$\\
      $\etaa$ & air viscosity (Gk)  & $[G][T]$ & $\equiv \etas$ & small air viscosity $\etaa/G\tau$\\
         $L_z$ & channel length & $[L]$ & $20$ & large aspect ratio $L_z/L_y$\\
      $\Lambda$ & midpoint sample length& $[L]$ & $16$ & large air gap $(L_z-\Lambda)/L_y$\\

            $\ell$ & polymer microscopic length & $[L]$ & 0.01 & small microscopic length $l/L_y$\\
      $\lm$ & air-polymer interface width & $[L]$ & 0.01 & small microscopic length $\lm/L_y$\\
      $M$ & molecular mobility & $[L]^2[G]^{-1}[T]^{-1}$ & 0.0001 & rapid phase equilibration\\
      \hline
      $\delta_y$ & numerical mesh size & $[L]$ & small & converge $\delta_y / L_y$ until no dependence\\
       $\delta_z$ & numerical mesh size & $[L]$ & small & converge $\delta_z / L_z$ until no dependence\\
        $\delta_t$ & numerical timestep & $[T]$ & small & converge $\delta_t / \tau$ until no dependence\\
        \hline
    \end{tabular}
  }
  \caption{Parameters; their dimensions in  modulus $[G]$, length $[L]$ and time $[T]$; values used in our  simulations; and notes. The first three parameters specify our choice of units. The second five are the key physical parameters to be varied in our study (four within each constitutive model). $a$ and $\alpha$ set the dependences of $\sigma$ and $N_2$ on $\gdot$, and have been explored in earlier works~\cite{hemingway2019edge}. The set from $\etaa$ to $M$ do not affect the key physics, provided each assumes an appropriately large or small value; the final set are numerical parameters, which we ensure are converged to the appropriate small limit. Parameter values used in the simulations are as given in this table unless explicitly described otherwise. Abbreviations: Johnson-Segalman (JS) and Giesekus (Gk).
}
  \label{tbl:params}
\end{table*}

In performing any simulation, we first take a rectangular slab of
fluid centred horizontally in the simulation box of
Fig.~\ref{fig:geometry} (which, recall, shows only the left half of
the area simulated), with length $\Lambda$ in the vorticity
direction. We then equilibrate the fluid-air coexistence in the
absence of shear, to the contact angle just described. We then add a
small perturbation to the interface's position $h(y)$ along the $z$
direction, $h(y)\to h(y)+10^{-8}\cos(\pi y/L_y)$. This is needed only
when the interface is vertical, $\theta=90^\circ$, but we add it for
all values of $\theta$, for consistency. The final steady state is 
independent of the small amplitude of the initial perturbation. 
It is also relatively robust against the shape of the added perturbation.
In Sec. \ref{sec:results1},
the viscoelastic stress is initialised to lie on the stationary homogeneous constitutive curve
for the shear rate in question, then we shear the fluid at that rate.
In Sec. \ref{sec:results2}, unless otherwise stated, we initialise with the corresponding 1D flow
state which may be banded or homogeneous depending on the shear-rate. In 
cases where the homogeneous state is metastable, we ensure that the initial state is 
banded by sweeping the shear-rate from an appropriate starting value.
We define the steady state interface displacement 
\begin{equation}
d = \max(h(y)) - \min(h(y)),
 \label{eq:displacement_def}
\end{equation}
with $d_0$ the value of this in a system without shear, noting that for $\theta = 90\degree$, $d_0 = 0$.

The parameters pertaining to the model equations, flow geometry, boundary conditions and imposed flow are summarised in Table.~\ref{tbl:params}, along with the values to be used in our simulations. Details of our numerical methods can be found in Ref.~\cite{hemingway2019edge}. In order to accurately reproduce the physics of the air-fluid coexistence, we require the air viscosity to be much smaller than the zero shear viscosity of the fluid (solvent and polymer combined), giving the condition $\etaa \ll \etas + G \tau$. In the first part of our study (Sec.~\ref{sec:results1}),  we further require each constitutive model to have a monotonic constitutive curve, to rule out true bulk shear banding in the absence of any edge effects. For the Giesekus model, it is possible to obtain monotonic constitutive curves for values of $\etas \ll G \tau$. The condition $\etaa \ll \etas + G \tau$ can then be met by adopting same air/solvent viscosity across the whole domain, $\etaa \equiv \etas$, with is also numerically convenient. Accordingly, we follow this strategy in our simulations of the Giesekus model. For the Johnson-Segalman model, solvent viscosities $\etas$ as large as 0.1 are needed to study marginally banded flows (Sec.~\ref{sec:results2}), and we can no longer equate $\etaa$ and $\etas$ while also satisfying the condition $\etaa \ll \etas + G \tau$. Therefore in this case we must simulate distinct air and solvent viscosities. To do so, we follow the method as described in detail in Ref.~\cite{hemingway2019edge}.

The numerical mesh size and timestep have maximal values of $\delta_y = \delta_z = 0.0052$, $\delta_t = 0.006$.
For each figure we have checked that the data are converged with respect to any 
further reduction of these numerical parameters. 

In the phase diagrams of Figs. \ref{fig:Gamma_n}, \ref{fig:Gamma_gdot}, \ref{fig:stress_hetero}, \ref{fig:PD_d_eta0.100}, \ref{fig:phase_diag_dob_eta0.100_b} and \ref{fig:PD_eta0.025} that follow below, we display the numerically obtained threshold for edge instability as a black dashed line. This is calculated as follows. In the early time dynamics of the edge instability in any simulation run, the edge perturbation grows as $\sim e^{\omega t}$ (or decays, in the stable regime). By tracking this early time exponential growth (or decay), for a simulation at any coordinate pair of parameter values in any phase diagram, we extract the growth rate, $\omega$, at that location in the phase diagram. By measuring this for a number of values of $\Gamma / G L_y$ on either side of the threshold (at any fixed $\etas - \eta_c$ in Fig. \ref{fig:Gamma_n}, for example), we then obtain the point where $\omega = 0$ by linear interpolation. This black dashed line is thus determined independently from the information shown by the symbols in the phase diagrams, which pertain to the nonlinear states obtained in the limit of long time $t \to \infty$ in any simulation run.

\section{Results: apparent bulk shear banding induced by precursors to edge fracture}
\label{sec:results1}

\begin{figure}[!t]
  \includegraphics[width=0.9\columnwidth]{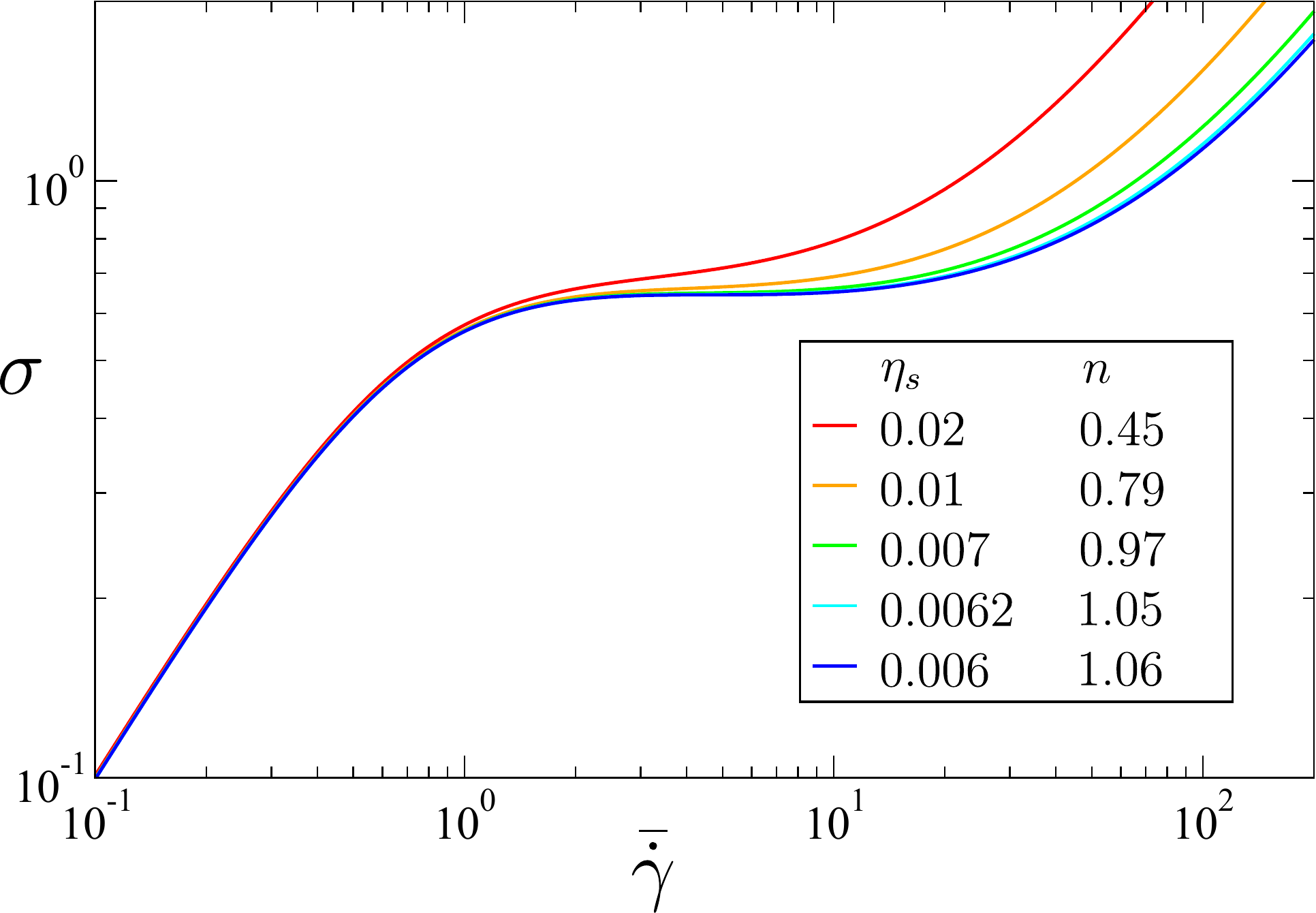}
  \caption{Constitutive curves of shear stress as a function of shear
  rate for stationary homogeneous shear flow. All curves are computed
  in the Giesekus model for a fixed value of the anisotropy parameter,
  $\alpha=0.8$. All curves are monotone increasing, but increasingly
  flatter curves downward pertain to increasingly more shear thinning
  fluids, with Newtonian viscosities $\etas = 0.02$, $0.01$, $0.007$,
  $0.0062$ and $0.006$. The caption also shows the equivalent plateau width $n$ 
  (see \eqref{eq:plateau_width_def} for definition).
}  \label{fig:constitutive}
\end{figure}

In this section, we explore in more detail the basic phenomenon
announced in a recent Letter~\cite{hemingway2018edge}: that even
relatively modest precursors to an edge fracture instability of the
fluid-air interface can induce apparent shear banding invading far
into the bulk of a strongly shear thinning fluid. Importantly, this is
true even if the underlying constitutive curve is monotone increasing,
$d\sigma/d\gdot > 0$, precluding true bulk banding in the absence of
edge effects. Accordingly, we restrict our attention throughout this
section to fluids for which the constitutive curve is indeed monotone
increasing. As we shall discuss, the apparent shear banding 
invades progressively further into the fluid bulk for progressively
flatter (but still monotone increasing) constitutive curves.

A set of constitutive curves computed within the Giesekus model for a
fixed value of the anisotropy parameter $\alpha$ is shown in
Fig.~\ref{fig:constitutive}. (We fix $\alpha=0.8$ throughout this
section.)  Although each curve is indeed monotone increasing, each has
a relatively flat quasi-plateau centred on a strain rate
$\gdotbar\approx
\sqrt{10}$. The flatness of this plateau will prove an important quantity in what follows. We shall quantify it via the plateau width
\begin{equation}
	n=\log_{10}(\gdot_{\rm h}/\gdot_{\rm l}),
\label{eq:plateau_width_def}
\end{equation}
determined by the shear-rates
$\gdot_h$, $\gdot_l$ at the extrema of the plateau, corresponding to
$\pm 5\%$ of the stress at the flattest point. In what follows, we
shall report results both in terms of the plateau width, $n$, and the
solvent viscosity, $\etas -\eta_{\rm c}$ (where $\eta_{\rm c}=0.005918$
is the value of $\etas$ below which the constitutive curve is
non-monotonic, for $\alpha=0.8$). We note that $n$ is directly set by
$\etas$, and is more directly measurable in any experiment. Stronger
shear thinning corresponds to lower values of $\etas$ and larger values
of $n$.

\begin{figure}[!t]
  \includegraphics[width=\columnwidth]{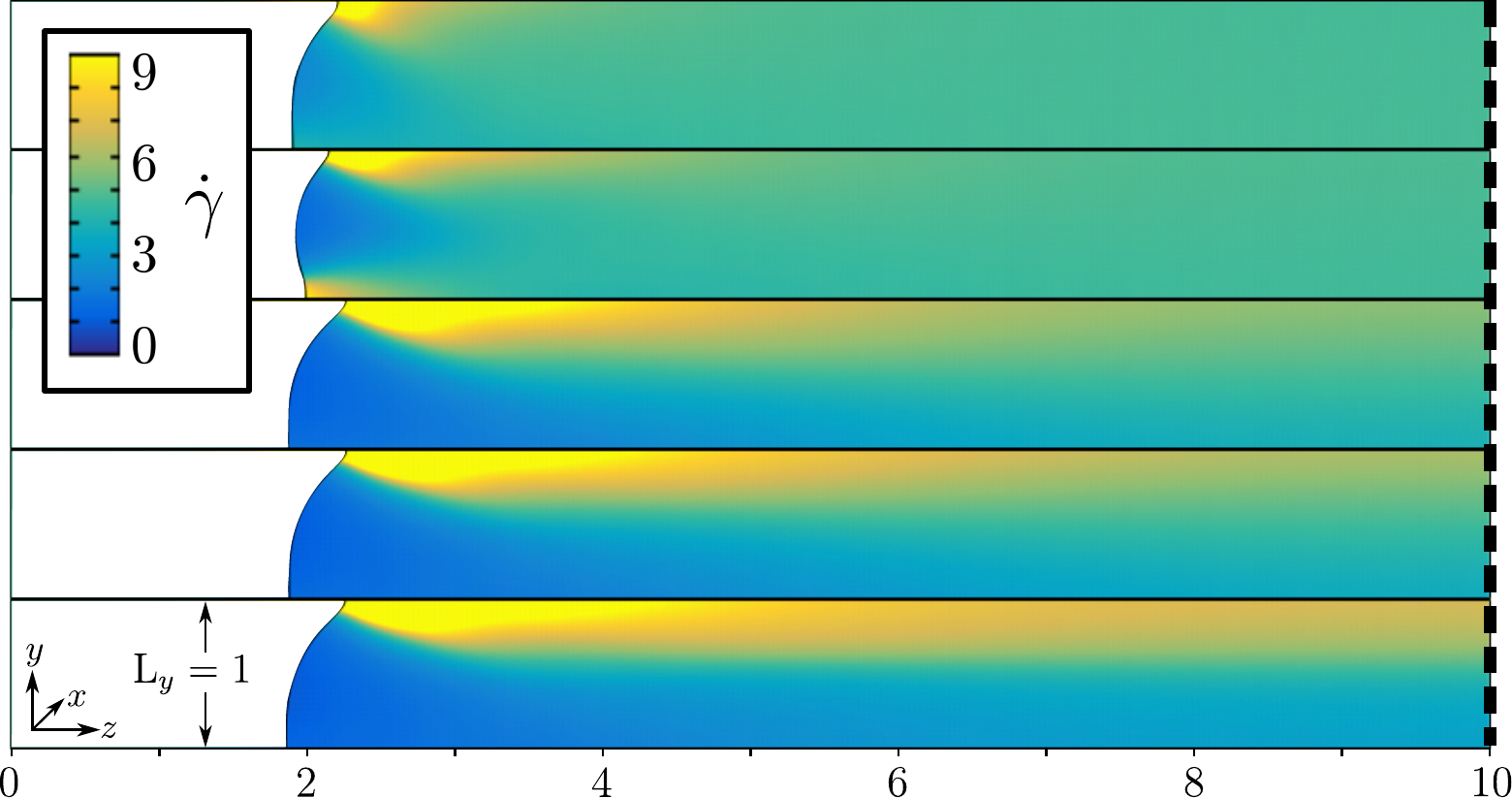}
  \caption{Simulation snapshots showing the  frame invariant local shear rate,
$\gdot(\vecv{r})=\sqrt{2\tens{D}(\vecv{r}):\tens{D}(\vecv{r})}$, at an imposed shear rate  $\gdotbar=4.7$, in the quasi-plateau regime of the constitutive curve, for a fluid-air interfacial tension $\Gamma = 0.16$. Snapshots downwards correspond to constitutive curves downwards in Fig.~\ref{fig:constitutive}. An apparent shear banding phenomenon invades  further into the fluid bulk for progressively flatter constitutive curves. $\theta = 90\degree$.}
  \label{fig:snapshots}
\end{figure}

\begin{figure*}[t]
  \includegraphics[width=0.75\textwidth]{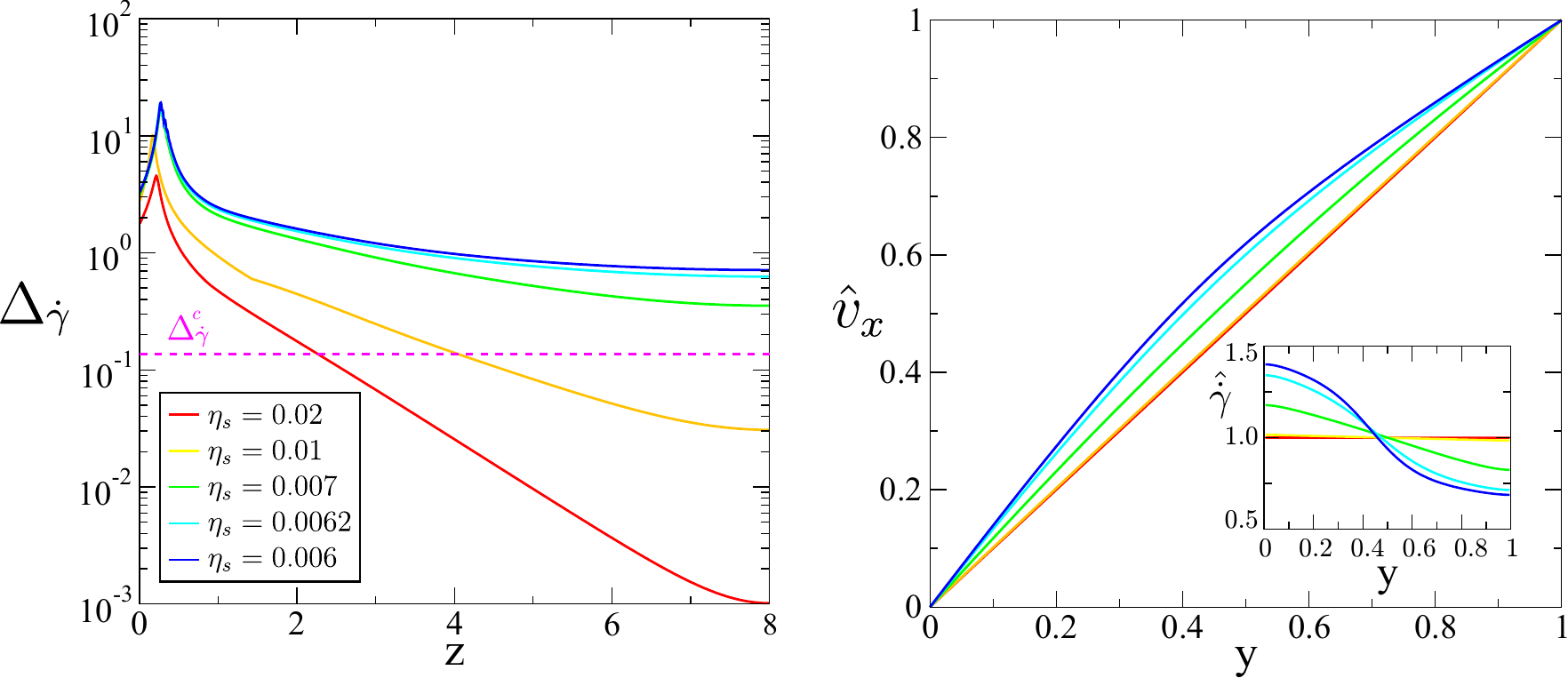}
  \caption{{\bf Left:} degree of apparent shear banding across the
  flow-gradient direction $y$ as a function of distance $z$ along the
  vorticity direction away from the fluid-air
  interface, into the bulk. Curves upwards correspond to increasingly flatter
  constitutive curves in Fig.~\ref{fig:constitutive}. Dashed magenta line
  indicates the value of the degree of banding above which banding is
  indeed obviously apparent in the velocity profiles. {\bf
  Right:} normalised velocity profiles $\hat{v}_x(y)$ across the cell
  midpoint $z = 8$ (black dashed line in
  Fig.~\ref{fig:snapshots}). Increasingly shear banded profiles
  correspond to  increasingly flatter constitutive curves in
  Fig.~\ref{fig:constitutive}. Inset: normalised local shear rate
  $\hat{\gdot}(y)$. $\gdotbar=4.7$, $\Gamma = 0.16$, $\theta = 90\degree$.} 
\label{fig:invasion}
\end{figure*}

Having discussed the homogeneous (0D) bulk constitutive curves, we now
present the results of our fully 2D simulations. Throughout this
section, we fix the value of the wetting angle, $\theta=90^\circ$.
Important parameters to be varied are then the solvent viscosity, $\etas$, 
which sets the shape of the constitutive curve in the way that we have
just discussed, the imposed shear rate, $\gdotbar$, and the surface
tension of the fluid-air interface, $\Gamma$. We focus on values of
the surface tension for which the fluid-air interface is modestly
disturbed by flow, giving partial edge fracture, as in the simulation
snapshots of Fig.~\ref{fig:snapshots}. In particular, we avoid the
regime of full edge fracture in which the interface loses its
integrity altogether and the fluid completely dewets the wall.
(Typical values of $\Gamma=\Gamma/GL_y$ are $0.001-0.1$ for synthetic
polymers and $0.1-10.0$ for DNA
solutions~\cite{Dimitriou2012,Schweizer2007,Ricci1984,Boukany2009,Hu2007,Hu2010,Skorski2011}.)

\begin{figure}[!t]
  \includegraphics[width=\columnwidth]{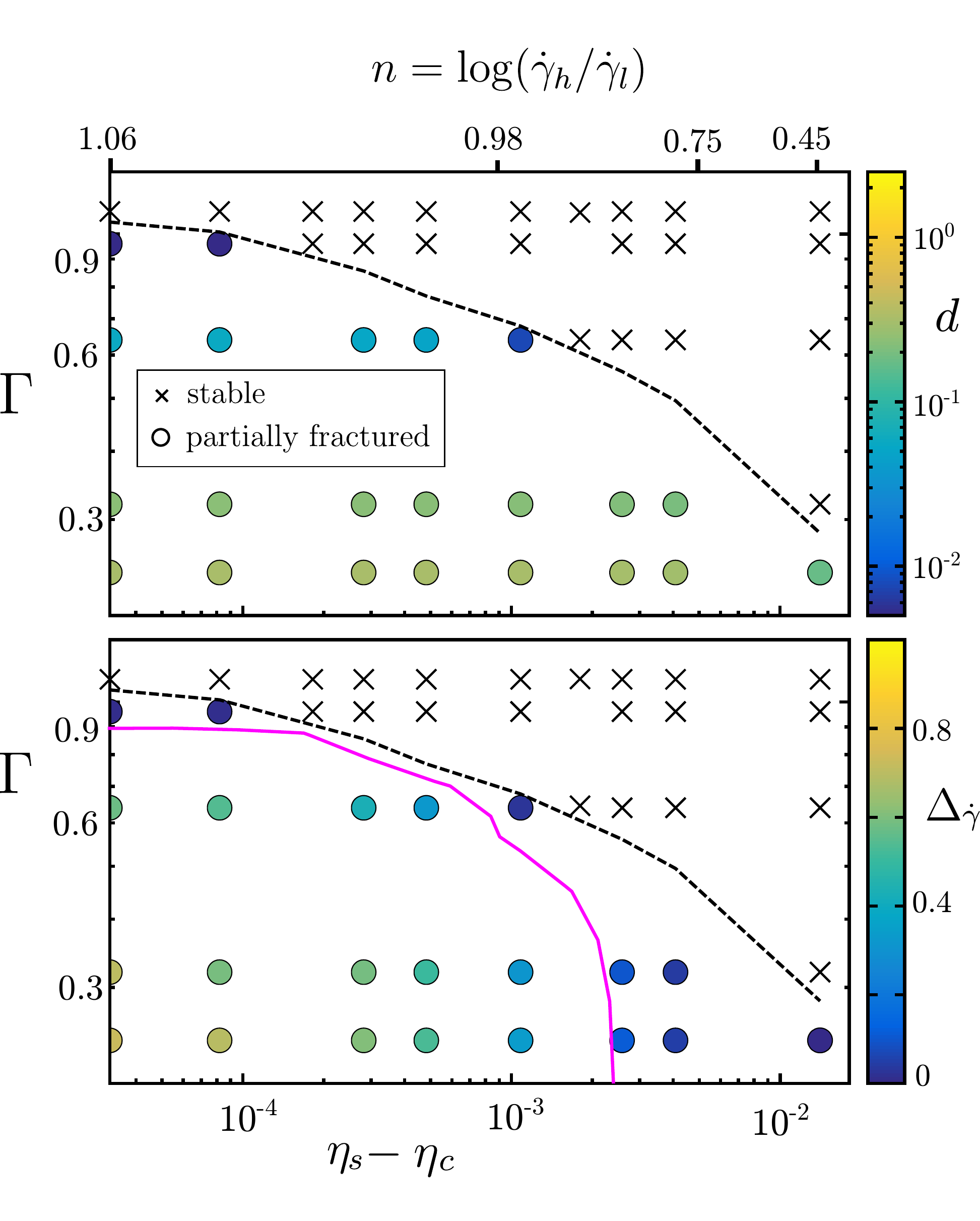}
  \caption{{\bf Top:} phase diagram showing the degree of disturbance of the
  air-fluid interface, $d$ (as defined in \eqref{eq:displacement_def}). Black dashed line: onset of positive
  eigenvalue for edge disturbance, as determined numerically. {\bf Bottom:} phase diagram of the
  degree of shear banding (as defined in \eqref{eq:degree_banding_def}) at the cell midpoint, $\dob(z=8.0)$. Magenta
  solid line: contour $\dobc = 0.15$, which we take as the threshold
  for visually apparent banding, as obtained from a linear interpolation of the data.  Both phase diagrams are shown in the
  plane of surface tension (ordinate) and an abscissa characterising
  the degree of of shear thinning in the constitutive curve. The
  bottom abscissa shows the value of the Newtonian viscosity, relative
  to that at which the constitutive curve develops a slope of zero,
  $\etas - \eta_c$. This decreases with increasing shear thinning right
  to left. The top abscissa shows the number of decades spanned by the
  quasi-plateau in the constitutive curve, $n$. This increases with
  increasing shear thinning right to left. Imposed shear rate $\gdotb
  = 4.7$, $\theta = 90\degree$.
}  \label{fig:Gamma_n}
\end{figure}

\begin{figure}[!t]
  \centering %
  \includegraphics[width=\columnwidth]{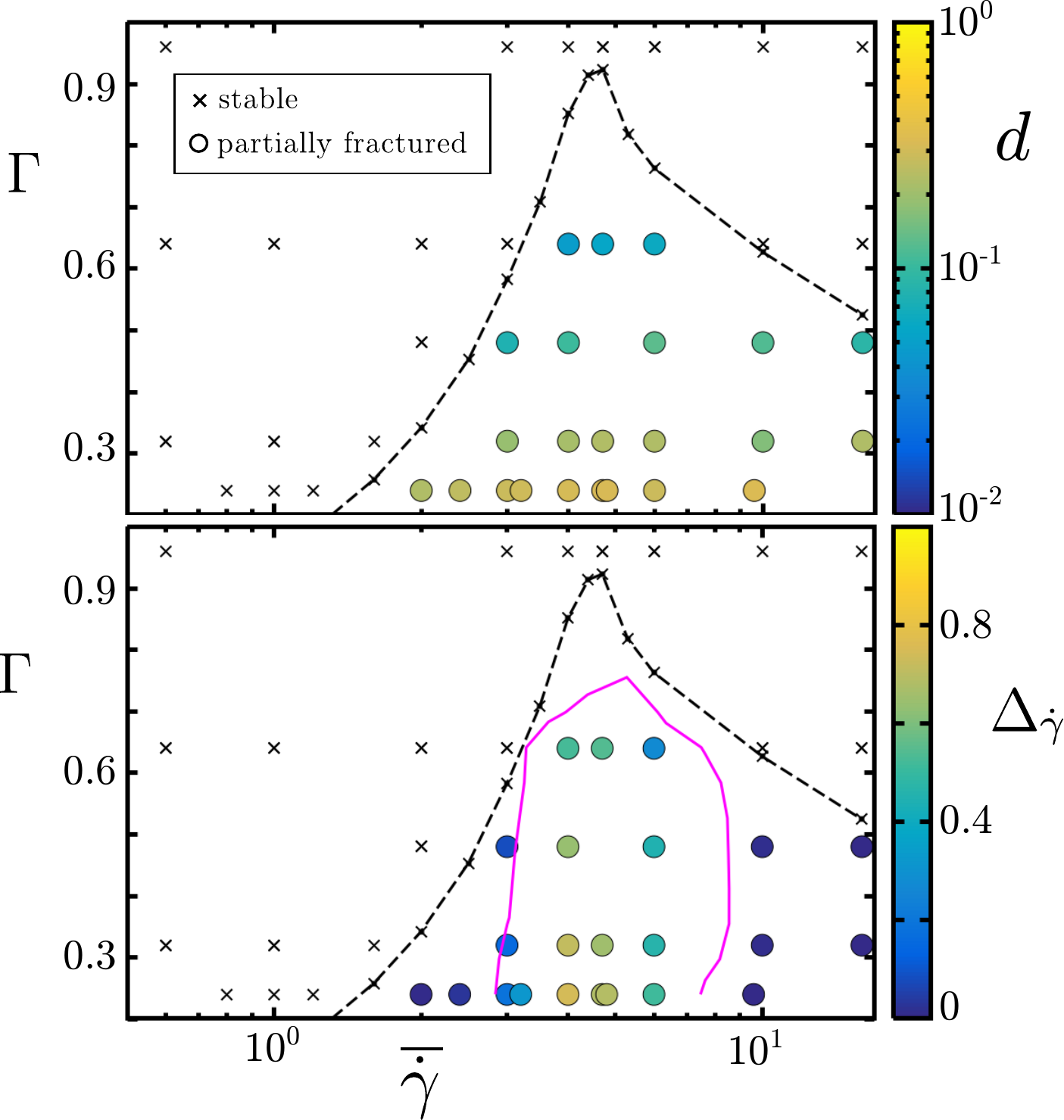}
  \caption{{\bf Top:} phase diagram of the degree of disturbance of the  air-fluid interface, $d$ (as defined in \eqref{eq:displacement_def}).  {\bf Bottom:} phase diagram of the degree of shear banding (as defined in \eqref{eq:degree_banding_def}) at the cell midpoint, $\dob(z=8.0)$. Magenta solid line: contour $\dobc = 0.15$, as obtained from a linear interpolation   of the data, which we take as the threshold  for visually observable shear banding. In both diagrams, the numerically determined onset of the positive eigenvalue for edge disturbance is shown (black dashed line).   Both phase diagrams are shown in the plane of surface tension (ordinate) and imposed shear rate (abscissa). Both pertain to a constitutive curve with a degree of shear thinning prescribed by a fixed value of Newtonian viscosity $\etas=0.006$, with a quasi-plateau spanning $n=1.06$ decades. $\theta = 90\degree$.} 
  \label{fig:Gamma_gdot}
\end{figure}

Although such modest edge disturbances may in themselves go unnoticed
experimentally, they are nonetheless capable of causing a much more
dramatic quasi-bulk shear banding phenomenon. This is increasingly
evident in successive snapshots downwards in Fig.~\ref{fig:snapshots},
which correspond to simulations performed for successive constitutive
curves downwards in Fig.~\ref{fig:constitutive}, at a fixed shear rate
$\gdotbar=4.7$ in the quasi-plateau regime in each case, for a surface
tension $\Gamma=0.16$. Indeed, an apparent shear banding phenomenon
that is strongly localised near the fluid-air interface for a
relatively steep constitutive curve invades progressively further
along the vorticity direction into the bulk, away from the fluid-air
interface, for progressively flatter constitutive curves.  The
corresponding velocity profile across the flow-gradient direction $y$
at the cell centre, $z=8$ (Fig.~\ref{fig:invasion}, right) accordingly
becomes increasingly banded. (We denote by $z=0$ the position of the
left hand fluid-air interface at the start of the simulation run.)

To quantify the degree to which the flow is shear banded across the
flow-gradient direction $y$ at any distance $z$ away from the fluid-air
interface, we define 
\begin{equation}
\dob(z) = \left[\gdot_\textrm{max}(z)-\gdot_\textrm{min}(z)\right]/\gdotb,
\label{eq:degree_banding_def}
\end{equation}
with $\gdot_\textrm{max}(z)$
the maximum shear rate at any point across $y$ at fixed $z$, and
$\gdot_\textrm{min}$ the counterpart minimum value. $\gdotbar$ is the
gap-averaged shear rate. By inspecting many profiles, we adopt $\dobc
= 0.15$ as the minimum threshold value of this measure to give
visually apparent banding in the velocity profile.  Plots of $\dob$ as
a function of distance $z$ into the bulk are shown in
Fig.~\ref{fig:invasion} (left). As can be seen, visually apparent
banding persists right to the centre of the flow cell for the flattest
constitutive curves in Fig.~\ref{fig:constitutive}. 

\begin{figure}[!t]
  \centering 
  \includegraphics[width=\columnwidth]{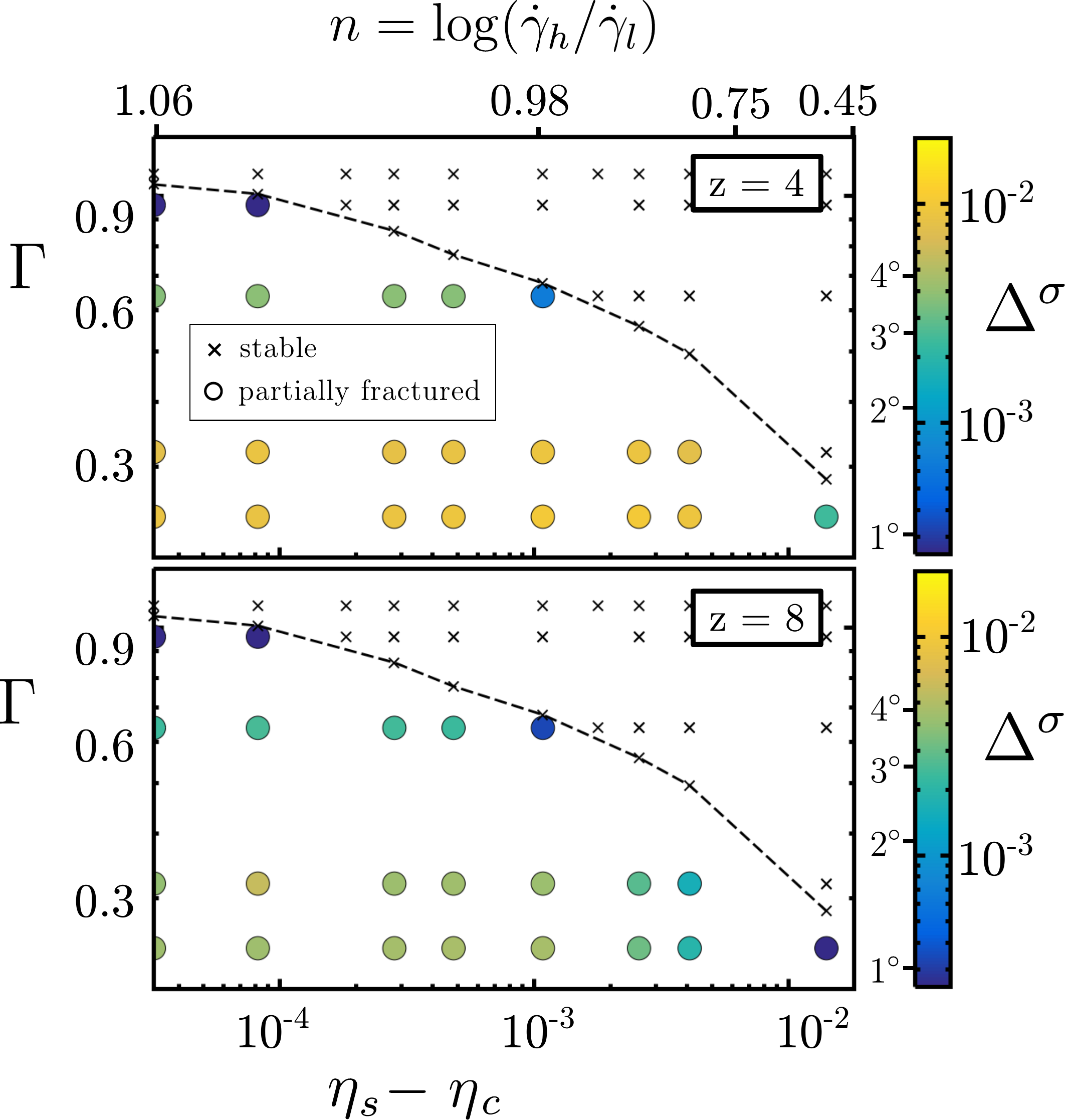}
  \caption{Phase diagrams showing the degree of shear stress heterogeneity
  $\dos = \sigma_\textrm{max} / \sigma_\textrm{min} - 1$ that arises due to the disturbance of the fluid-air interface at a distance of $z = 4.0$ (top) and $8.0$ (bottom) into the fluid bulk. For reference, the colourbars are additionally marked with the degree of stress heterogeneity that would arise in a cone and plate device with a cone angle of $1, 2, 3$ or  $4\degree$.  Imposed shear rate $\gdotb = 4.7$, $\theta = 90\degree$.} \label{fig:stress_hetero}
\end{figure}

So far, we have shown that the degree to which disturbances at the
fluid-air interface can induce apparent shear banding into the fluid
bulk increases with increasing shear thinning in the underlying
constitutive curve.  In doing so, we have kept the surface tension
$\Gamma$ of the fluid-air interface fixed, and considered a single
value of the imposed shear rate, $\gdotbar$, near the flattest part of
the constitutive curve. We now explore the effects on this phenomenon
of also (first) varying the surface tension, $\Gamma$, and then
(second) the imposed shear rate, $\gdotbar$.

Phase diagrams are shown in Fig.~\ref{fig:Gamma_n} of the degree of
disturbance of the fluid-air interface (top) and of apparent shear
banding at the midpoint of the fluid bulk (bottom) in the plane of
surface tension, $\Gamma$, and degree of shear thinning in the
constitutive curve, $n$, again for a fixed value of the imposed shear
rate, $\gdotbar$, near the flattest part of the constitutive curve. A
horizontal slice across this phase diagram accordingly spans a suite of
constitutive curves with varying degrees of shear thinning, with lower
values of $n$ right to left giving stronger shear thinning. For a
fixed value of surface tension, the degree of edge disturbance and of
shear banding at the cell midpoint both increase with increasing shear
thinning from right to left. For a fixed degree of shear thinning, the
same quantities increase with decreasing surface tension, from top to
bottom.

Phase diagrams of the same quantities are again shown in
Fig.~\ref{fig:Gamma_gdot} but now in the plane of surface tension,
$\Gamma$, and imposed shear rate, $\gdotbar$, for a fixed degree of
shear thinning in the constitutive curve, $n$. A horizontal slice
across this phase diagram accordingly spans a range of imposed shear rates
for one particular constitutive curve with a fixed degree of shear
thinning. For a fixed value of surface tension less than about $0.8$,
the degree of edge disturbance and of shear banding at the cell
midpoint are noticeable for values of the shear rate in the
quasi-plateau regime of the constitutive curve. For any fixed value of
shear rate on this quasi-plateau, the same quantities increase with
decreasing surface tension of the fluid-air interface, top to bottom
down the phase diagram.

How can we understand this apparent shear banding phenomenon? As
discussed in Ref.~\cite{hemingway2019edge}, the physical mechanism of
the instability that leads to significant disturbances of the
fluid-air interface involves a perturbation $\delta\Sigma_{xy}(y,z)$
to the shear stress field
across the flow gradient direction $y$, which gradually decays as a
function of the distance $z$ away from fluid-air interface. This leads
to a corresponding disturbance $\delta\gdot(y,z)$ in the shear rate
field, by an amount that scales as the inverse slope of the
constitutive curve. For a strongly shear thinning fluid, therefore,
even a relatively modest disturbance to the shear stress creates a
large disturbance of the shear rate, causing apparent shear banding.

A related question is then how these stress and shear rate
heterogeneities associated with disturbances of the fluid-air
interface compare with those associated with the geometries of typical
flow cells. This is especially pertinent in light of the results of 
Ref.~\cite{AdamsOlmsted2009} which demonstrated that stress gradients 
(originating from geometry) can cause shear banding phenomena in fluids 
that are described by a monotonic underlying constitutive curve.

Fig.~\ref{fig:stress_hetero} shows phase diagrams of the
degree of stress heterogeneity across the flow gradient direction $y$
caused by disturbances in the fluid-air interface, at distances of
$z=4.0$ (top) and $z=8.0$ (bottom) into the fluid. For reference, the 
colourbar additionally shows the level of stress heterogeneity that 
would instead arise from device geometry for cone-plate cells of various cone angles. 
Given that typical cone-angles are about $0.5-1^\circ$, the
stress heterogeneity and corresponding apparent shear banding that we
observe clearly exceed those arising from typical cell geometries.

\section{Edge fracture induced by shear banding}
\label{sec:results2}

\subsection{Marginally non-monotonic constitutive curve}
\label{sec:marginal}

\begin{figure}[!t]
  \includegraphics[width=\columnwidth]{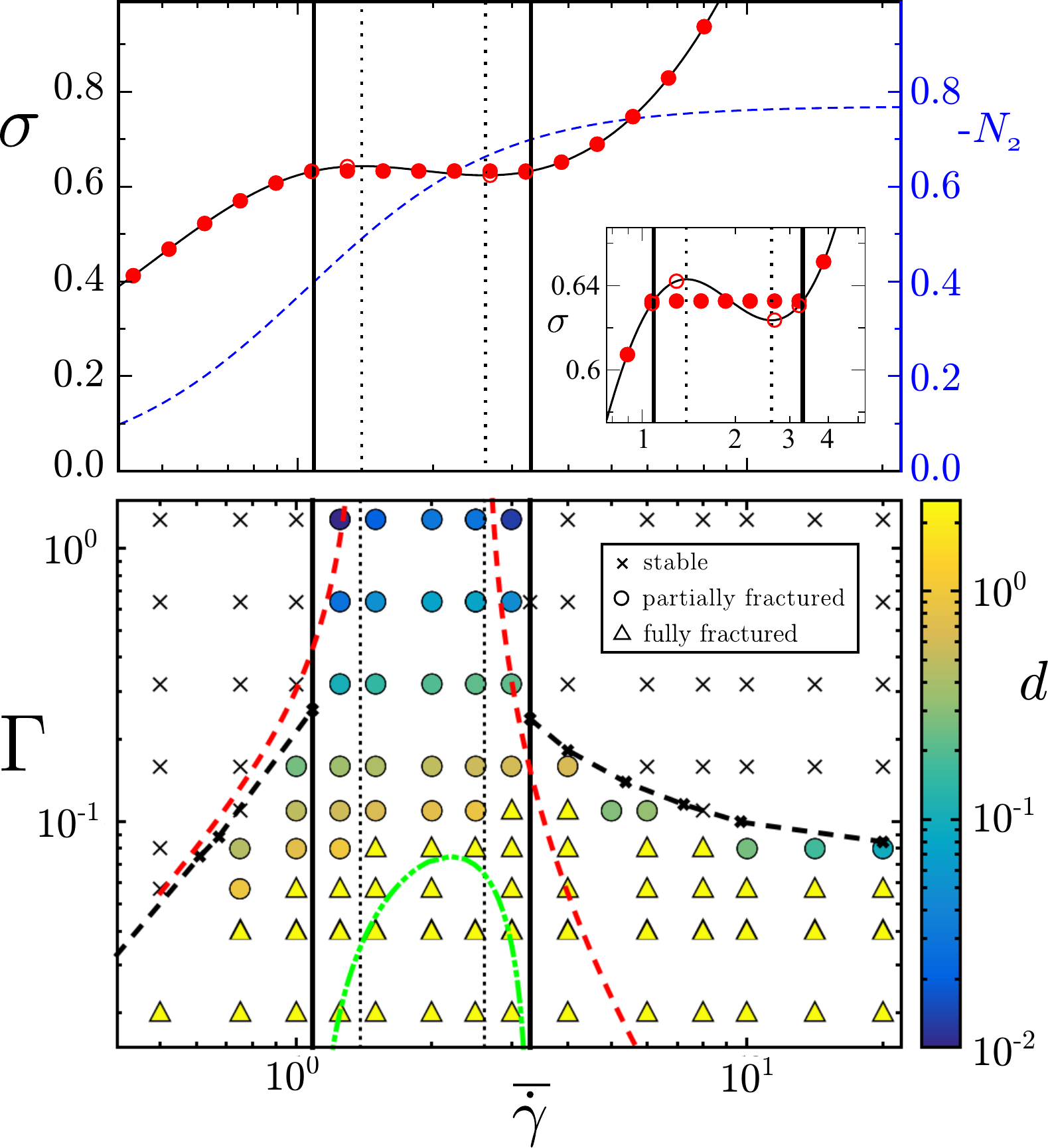}
  \caption{\textbf{Top:} Homogeneous (0D) constitutive curve (black
  curve) and steady state flow curve computed in a 1D simulation (red
  solid circles). Inset shows the same data, zoomed to focus on the
  shear banding regime. Open red circles denote metastable states of
  homogeneous shear flow. Dashed blue line shows (minus) the second
  normal stress, $-N_2$. \textbf{Bottom:} Colour map of degree of
  disturbance to the fluid-air interface, $d$, in the plane of the
  surface tension of that interface, $\Gamma$, and the imposed shear
  rate, $\gdotb$.    Black dashed lines denote the numerically determined onset of edge disturbance, red dashed line marks the threshold for instability as determined via linear stability analysis, and green dash-dotted  line shows the analytical prediction of Ref.~\cite{Skorski2011}.
  Johnson-Segalman model, $\etas = 0.1$, $\theta = 90\degree$.}
  \label{fig:PD_d_eta0.100}
\end{figure}

We consider now a fluid with a non-monotonic constitutive curve,
giving a bulk flow that is shear banded over some range of strain
rates (even in the absence of any edge effects). In particular, we
seek to appraise using our simulations the analytical predictions of
Skorski and Olmsted~\cite{Skorski2011}: that shear banding in the
fluid bulk can induce full edge fracture at the fluid-air interface,
for some values of the surface tension and imposed shear rate. It should
be noted that the work in Ref.~\cite{Skorski2011} did not perform
a dynamic calculation, nor did it consider the effect of secondary flows.
We start by considering a constitutive curve that is only marginally
non-monotonic, giving a relatively small jump in the shear rate across
the interface between the shear bands, before discussing in the next
subsection~\ref{sec:strongly} below a more highly non-monotonic curve,
giving a larger jump.

A marginally non-monotonic constitutive curve is accordingly shown in
Fig.~\ref{fig:PD_d_eta0.100} (top). The corresponding second normal
stress as a function of shear rate is shown on the right hand
axis. These two curves pertain to a state of stationary homogeneous
(0D) shear flow. In the regime of negative constitutive slope,
$d\sigma/d\gdot<0$,
 an initially homogeneous flow is linearly unstable to the formation
 of shear bands. The shear rates delineating this regime are shown by
 vertical dotted lines in Fig.~\ref{fig:PD_d_eta0.100}. The steady
 flow state, computed within a 1D calculation that now allows shear
 banding (but still ignores edge effects) is then shear banded. This
 leads to a plateau in the steady state flow curve (red circles). The
 shear rates delineating this regime are shown by vertical solid
 lines. Between the dotted and solid vertical lines, an initially
 homogeneous flow is metastable to the formation of shear bands.

\begin{figure}[!t]
  \includegraphics[width=\columnwidth]{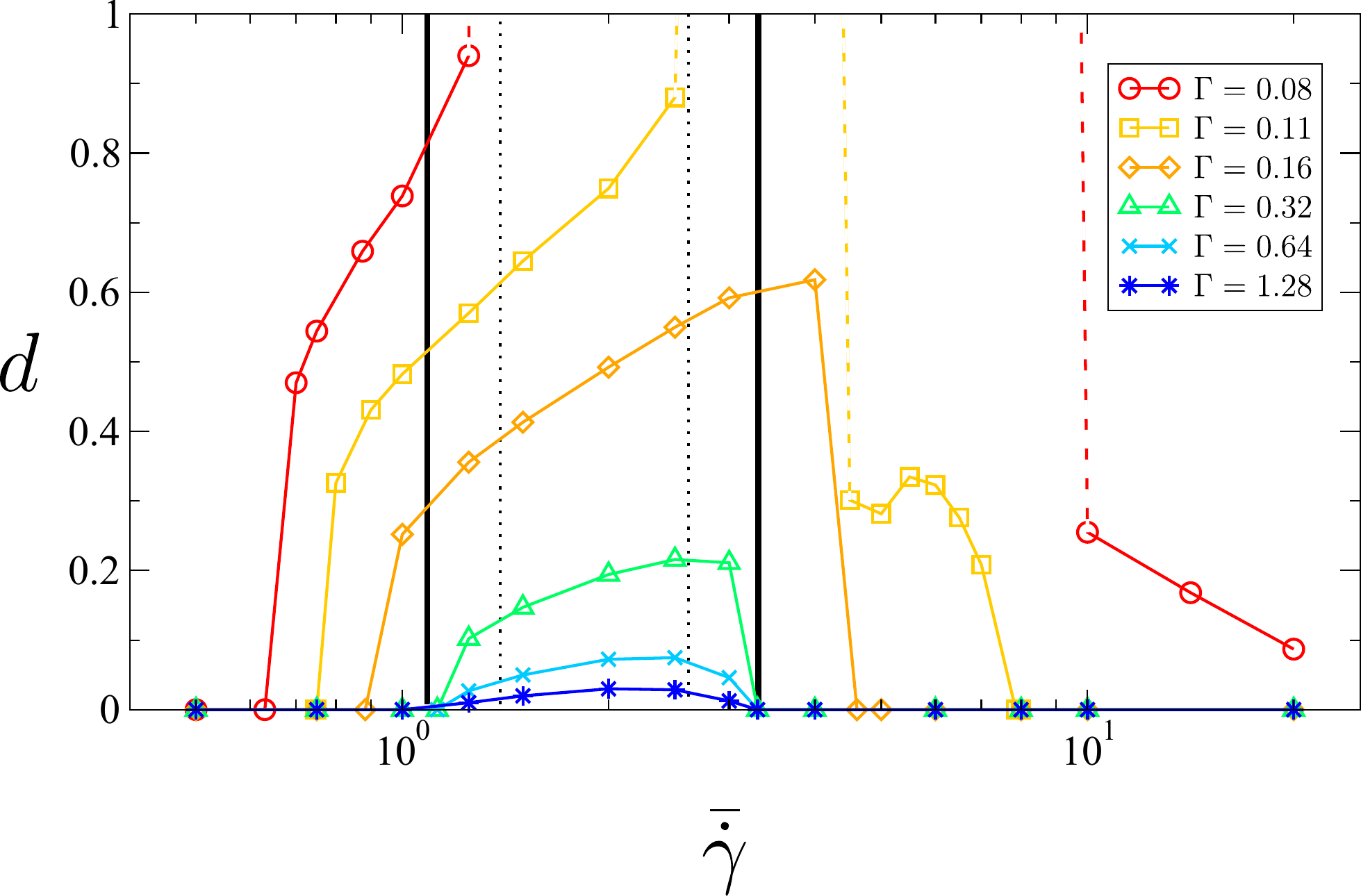}
  \caption{Horizonal slices across the colourmap of
  Fig.~\ref{fig:PD_d_eta0.100} (bottom) showing the degree of
  disturbance to the fluid-air interface, $d$, as a function of the
  imposed shear rate, $\gdotbar$, for several values of the surface
  tension of the fluid-air interface, $\Gamma$.}
  \label{fig:disp_eta0.100_here}
\end{figure}

\begin{figure}[!b]
  \includegraphics[width=\columnwidth]{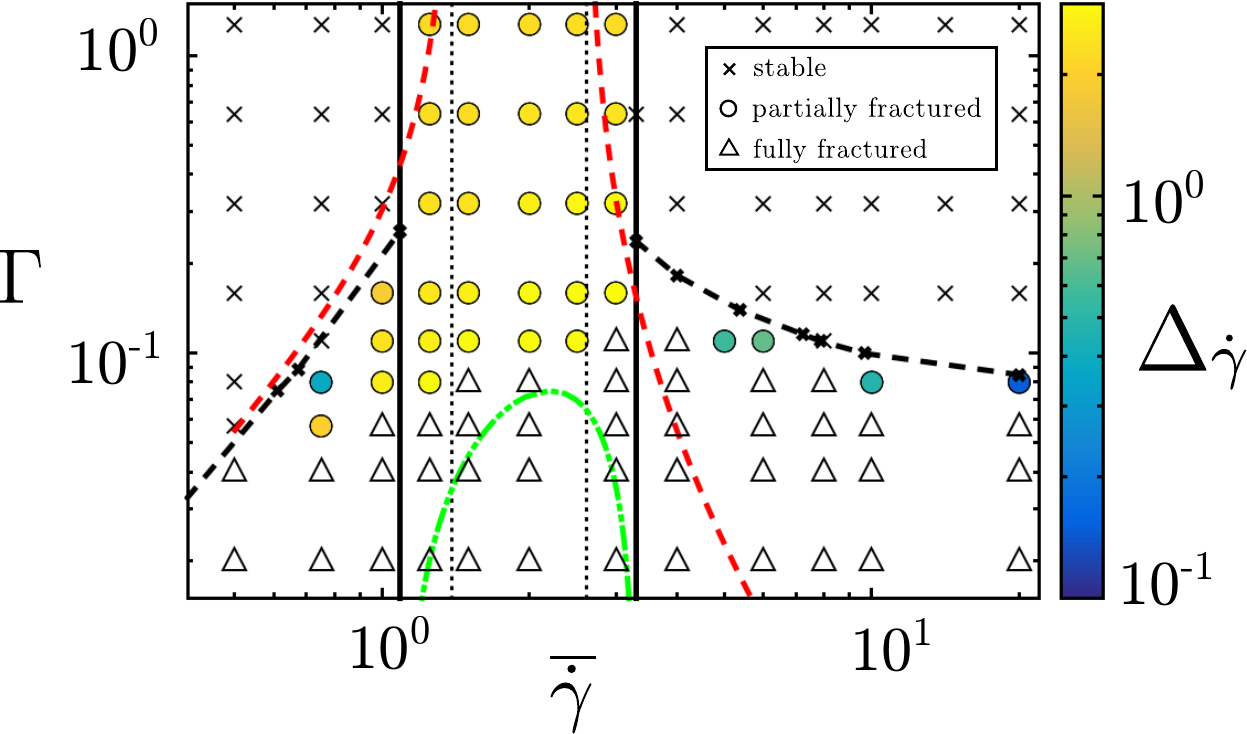}
  \caption{Colour map showing the degree of shear banding $\dob$ (as defined in \eqref{eq:degree_banding_def}),
  measured at a position $z=1$ into the bulk away from the fluid-air
  interface, shown in the same plane of surface tension and shear rate
  as in Fig.~\ref{fig:PD_d_eta0.100}. Open triangles denote fully edge
  fractured states, in which $\dob$ cannot be reliably measured.
  Black dashed lines denote the numerically determined
  onset of edge disturbance. Red dashed line marks the threshold for
  instability as determined via linear stability analysis. Green dash-dotted
  line shows the analytical prediction of Ref.~\cite{Skorski2011}.
  Johnson-Segalman model, $\etas = 0.1$, $\theta = 90\degree$.}
\label{fig:phase_diag_dob_eta0.100_b}
\end{figure}

\begin{figure*}[!t]
\centering
  \includegraphics[width=1.0\textwidth]{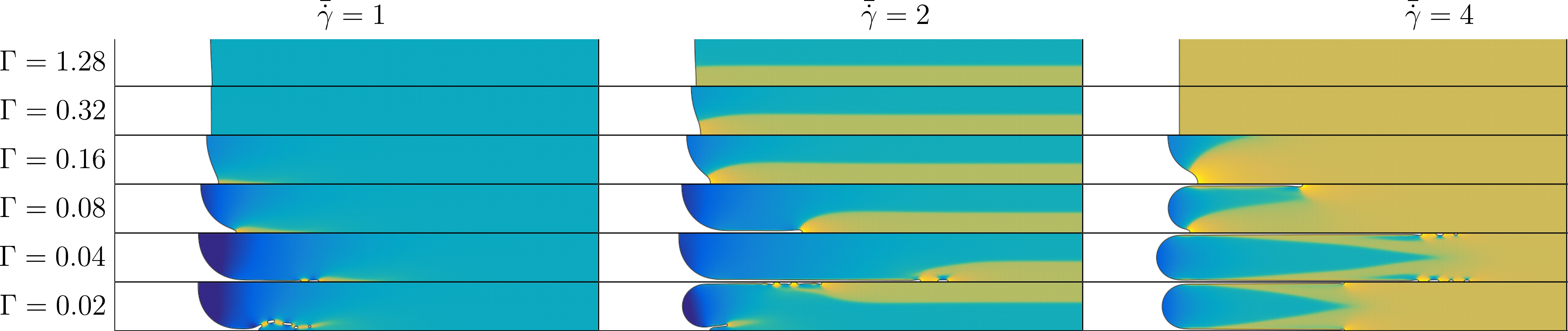}
  \caption{Simulation snapshots showing the locally invariant shear,
  rate $\gdot(\vecv{r})=\sqrt{2\tens{D}(\vecv{r}):\tens{D}(\vecv{r})}$,
  for the case of a marginally non-monotonic constitutive
  curve. Applied shear-rate $\gdotb = 1.0, 2.0, 4.0$ (left to
  right) and surface tension $\Gamma = 0.02, 0.04, 0.08, 0.16, 0.32,
  1.28$ (bottom to top). Johnson-Segalman model, $\etas =
  0.1$, $\theta = 90\degree$.}
  \label{fig:snapshots_eta0.100_b}
\end{figure*}

Having discussed these 0D and 1D states, we now discuss our fully 2D
simulations, in which a sheared slab of fluid coexists with the
outside air, separated by a fluid-air interface. These simulations
allow a study of any edge fracture-like disturbances at that fluid-air
interface, as well as any shear banding within the fluid bulk.

A colourmap showing the degree of disturbance of the fluid-air
interface as a function of the surface tension of that interface,
$\Gamma$, and the imposed shear rate, $\gdotbar$, can be found in
Fig.~\ref{fig:PD_d_eta0.100} (bottom). Outside the regime where the
bulk flow is shear banded, the interface remains undisturbed at high
values of the surface tension, but is always disturbed at lower values
of the surface tension. The red dashed lines show the analytical
prediction of Hemingway and
Fielding~\cite{Hemingway2017,hemingway2019edge} for the onset of a
linear instability to these disturbances -- {\it i.e.} for the onset
of partial edge fracture -- for this case of a homogeneous (unbanded)
bulk shear flow. This was derived in the limit of low strain rates,
where it indeed agrees well with the onset of partial edge fracture in
our simulations (black dashed lines).

Consider now imposed shear rates inside the shear banding regime,
within the vertical solid lines.  Here the fluid-air interface is
always disturbed to some extent, although by an amount that diminishes
with increasing surface tension. This is consistent with the work of
Skorski and Olmsted~\cite{Skorski2011}, who noted that an edge
disturbance is an inevitable consequence of force balance for a shear
banded flow, due to the jump in second normal stress across the
interface between the bands.

Having explored in Fig.~\ref{fig:PD_d_eta0.100} the degree of edge
disturbance, $d$, as a function of surface tension, $\Gamma$, and
imposed shear rate, $\gdotbar$, we show in
Fig.~\ref{fig:disp_eta0.100_here} a collection of horizontal slices
across this plane, now plotting $d$ as a function of $\gdotbar$ for
several fixed values of $\Gamma$. Consistent with the above
discussion, the interface is always disturbed to some degree inside
the shear banding regime. Outside the banding regime, the interface is
undisturbed for large values of surface tension, but disturbed for
smaller values.

A noticeable feature of Fig.~\ref{fig:disp_eta0.100_here} is that, for
several values of the surface tension, the degree of edge disturbance
varies smoothly across the critical shear rate that marks the onset of
bulk shear banding. This is a priori surprising: within a 1D
calculation, free of edge effects, the bulk flow undergoes a
qualitative transition from unbanded to banded across that critical
shear rate. One might therefore expect very different levels of edge
disturbance once the fluid-air interface is accounted for in 2D. The
resolution to this puzzle lies in recognising that the edge
disturbances actually in turn modify the bulk state, at least for some
distance away from the fluid-air interface into the fluid: as seen for
a shear rate $\gdot=1.0$ and surface tension values $\Gamma=0.08,0.16$
in Fig.~\ref{fig:snapshots_eta0.100_b}, the flow becomes quasi-banded
a little way into the bulk, even at a shear rate a little below that
for which a purely 1D calculation (without a fluid-air interface)
would predict banding. This is simply another (albeit relatively mild)
manifestation of the phenomenon discussed Sec.~\ref{sec:results1}
above. A colourmap of the degree to which the flow is shear banded a
little distance ($z=1$) into the fluid away from the fluid-air
interface indeed confirms this.  See
Fig.~\ref{fig:phase_diag_dob_eta0.100_b}.

\begin{figure}[!b]
  \includegraphics[width=\columnwidth]{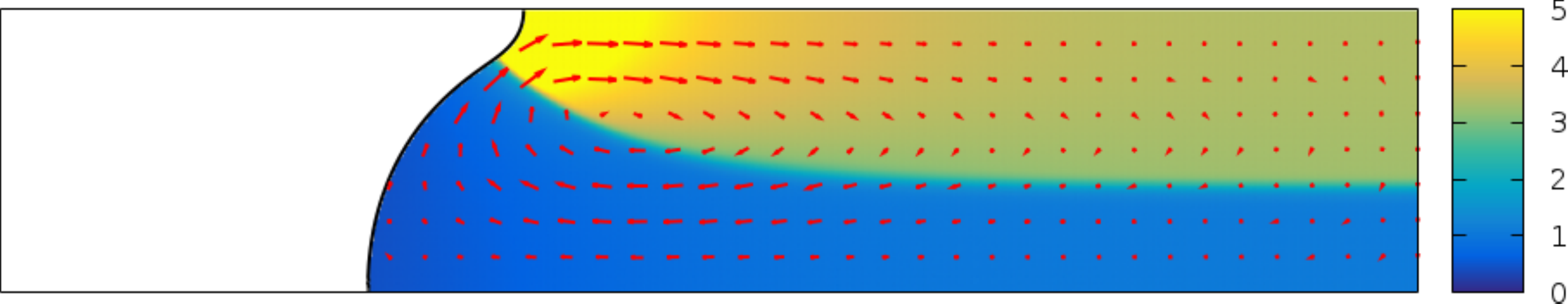}
  \caption{Colourmap of the frame invariant shear-rate,
  $\gdot(\vecv{r})=\sqrt{2\tens{D}(\vecv{r}):\tens{D}(\vecv{r})}$,
  showing secondary flows (red arrows) near the fluid-air
  interface. Johnson-Segalman model, $\gdotb = 2.5$, $\Gamma
  = 0.16$, $\etas = 0.1$, $L_z = 10$, $\Lambda = 7$, $\theta = 90\degree$.}
  \label{fig:secondary_flows}
\end{figure}

\begin{figure}[!t]
  \includegraphics[width=\columnwidth]{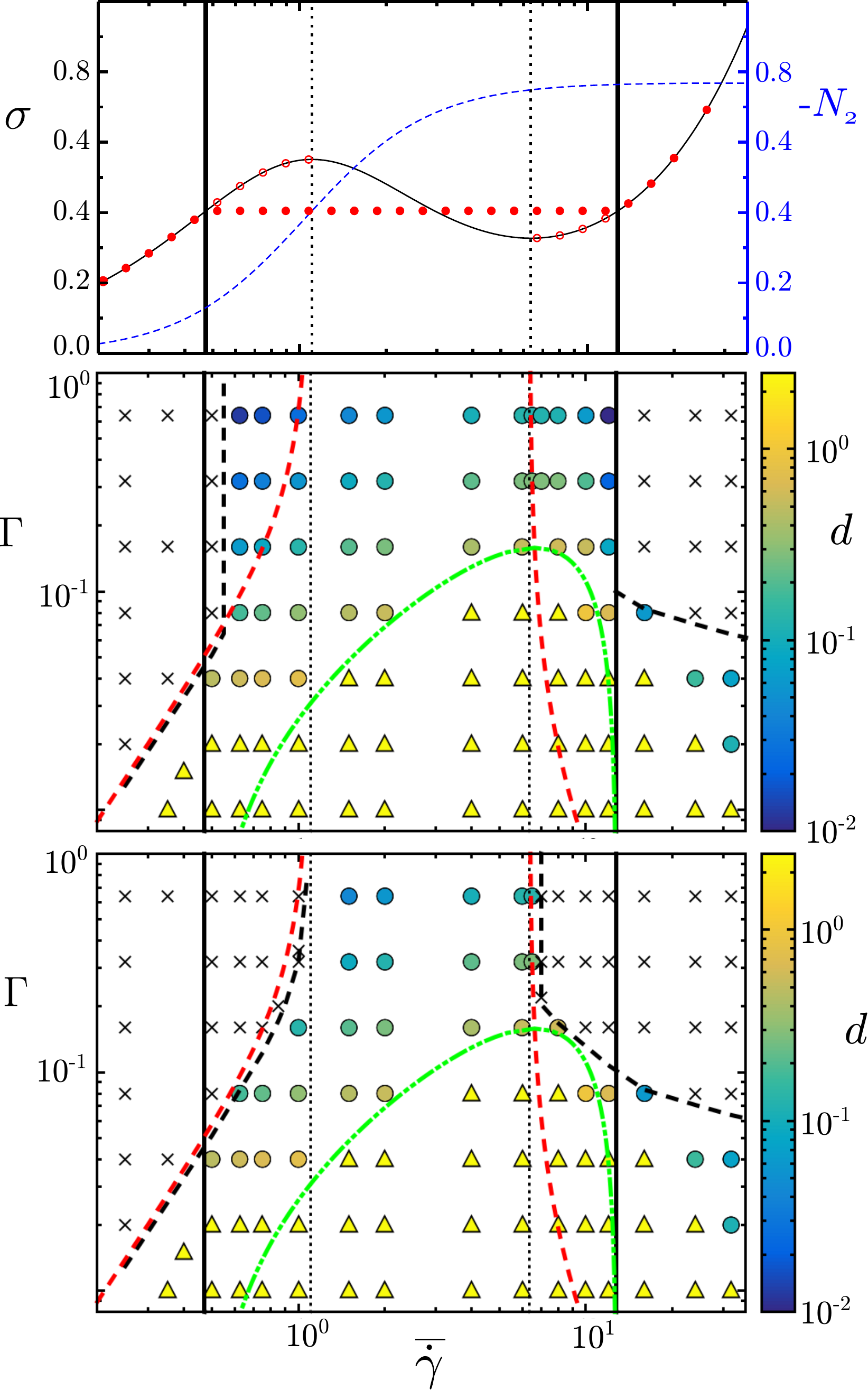}
  \caption{\textbf{Top:} Homogeneous (0D) constitutive curve (black
  curve) and steady state flow curve computed in a 1D simulation (red
  solid circles).  Open red circles denote metastable states of
  homogeneous shear flow. Dashed blue line shows (minus) the second
  normal stress, $-N_2$. \textbf{Middle:} Colour map of degree of
  disturbance to the fluid-air interface $d$ (as defined in \eqref{eq:displacement_def}), in the plane of the
  surface tension of that interface, $\Gamma$, and the imposed shear
  rate, $\gdotb$.  \textbf{Bottom:} Same as in middle figure, but for
  an initial condition of metastable homogeneous shear in the window
  of shear rates between the solid and dashed vertical lines. 
  Black dashed lines denote the numerically determined onset of edge disturbance, red dashed line marks the threshold for instability as determined via linear stability analysis, and green dash-dotted  line shows the analytical prediction of Ref.~\cite{Skorski2011}.
  Johnson-Segalman model, $\etas = 0.025$, $\theta = 90\degree$. }
   \label{fig:PD_eta0.025}
\end{figure}

As discussed above, at low values of surface tension the fluid-air
interface always becomes fully fractured. Skorksi and
Olmsted~\cite{Skorski2011} predicted the location of the transition
from partial to full fracture for a shear banded state by arguing
that, as the relative fraction of the rheometer gap taken up by each
band changes with the imposed shear rate, one of the bands can develop
a width that is unable to support the meniscus curvatures demanded by
the second normal stress balance, leading to full fracture.  Their
prediction is marked by a green dashed line in
Figs.~\ref{fig:PD_d_eta0.100}
and~\ref{fig:phase_diag_dob_eta0.100_b}. The regime in which our
simulations display full fracture is shown by yellow triangles. The
agreement between the analytical predictions and our simulations is at
best fair, possibly because Skorski and Olmsted assumed an initial
``base'' state with no secondary flows. Secondary flows nonetheless
inevitably arise, as seen in Fig.~\ref{fig:secondary_flows}.

\subsection{Highly non-monotonic constitutive curve}
\label{sec:strongly}

We consider finally a fluid with a highly non-monotonic constitutive
curve, giving a large jump in shear rate across the interface between
the bands. The top two panels of Fig.~\ref{fig:PD_eta0.025} show
results for this case in the same format as the two panels of
Fig.~\ref{fig:PD_d_eta0.100} for a marginally non-monotonic
curve. Many of the same features are apparent.  Outside the regime
where the bulk flow is shear banded, the fluid-air interface remains
undisturbed at high values of the surface tension, but is disturbed at
lower values. At low shear rates, the numerically observed transition
from an undisturbed to a disturbed fluid-air interface (black dashed
line) agrees well with the analytical prediction of Hemingway and
Fielding given a homogeneous (unbanded) bulk flow (red dashed
line)~\cite{Hemingway2017,hemingway2019edge}.

\begin{figure}[!t]
\centering
\includegraphics[width=1.0\columnwidth]{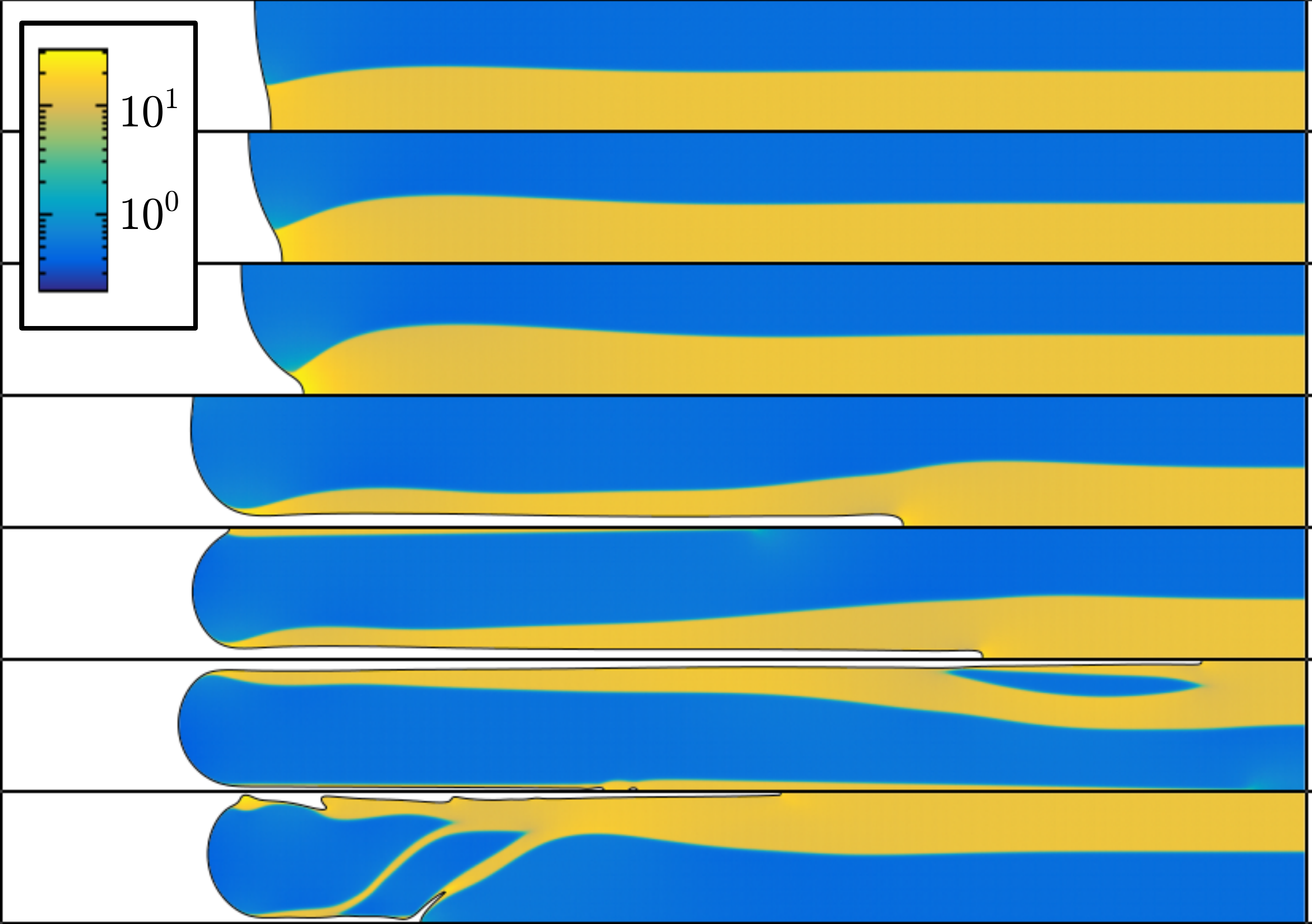}
\caption{ Simulation snapshots showing the locally invariant shear rate,
  $\gdot(\vecv{r})=\sqrt{2\tens{D}(\vecv{r}):\tens{D}(\vecv{r})}$, for
  the case of a highly non-monotonic constitutive curve. Applied
  shear-rate $\gdotb = 6.0$ with a surface tension $\Gamma = 0.01,
  0.02, 0.04, 0.08, 0.16, 0.32, 0.64$ (bottom to
  top). Johnson-Segalman model, $\etas = 0.025$, $\theta = 90\degree$.}
\label{fig:snapshots_eta0.025}
\end{figure}

In contrast, when the bulk flow is shear banded the fluid-air
interface is always disturbed to some extent, although by an amount
that diminishes with increasing surface tension. This is again
consistent with the argument of Skorski and Olmsted, that a shear
banded state must always have a disturbed interface on account of the
second normal stress jump across the interface between the
bands~\cite{Skorski2011}.  Snapshots of the locally invariant shear
rate for a fixed shear rate $\gdotbar=6.0$ inside the shear banding
regime are shown in Fig.~\ref{fig:snapshots_eta0.025}, with the
surface tension increasing in subpanels upward.

At low values of surface tension, the fluid-air interface always
becomes fully fractured, as shown by the yellow triangles in
Fig.~\ref{fig:PD_eta0.025}. Skorski and Olmsted's
prediction~\cite{Skorski2011} for the onset of full fracture in the
shear banding regime is marked by a green dashed line in
Fig.~\ref{fig:PD_eta0.025}. It shows fair agreement with our
simulations -- which is perhaps mainly fortuitous, given the poorer
agreement in Fig.~\ref{fig:PD_d_eta0.100} above, and (as we shall
shortly discuss) in Fig.~\ref{fig:PD_wetting} below.

In a narrow window of shear rates at the lower end of the regime where
shear banding is expected, the bulk flow is unexpectedly homogeneous,
and the fluid-air interface is accordingly unexpectedly
undisturbed. See the black crosses for an imposed shear rate
$\gdotbar=0.5$ and surface tension $\Gamma\ge 0.08$ in
Fig.~\ref{fig:PD_eta0.025} (middle). For such shear rates, the width
of the low shear band itself would be smaller than the finite
thickness of the interface itself. The flow therefore remains
unbanded. Simulations for a much smaller value of the parameter that
sets the thickness of the interface between the shear bands would be
shear banded, and would presumably then also have a disturbed
fluid-air interface.

\begin{figure}[!t]
  \includegraphics[width=\columnwidth]{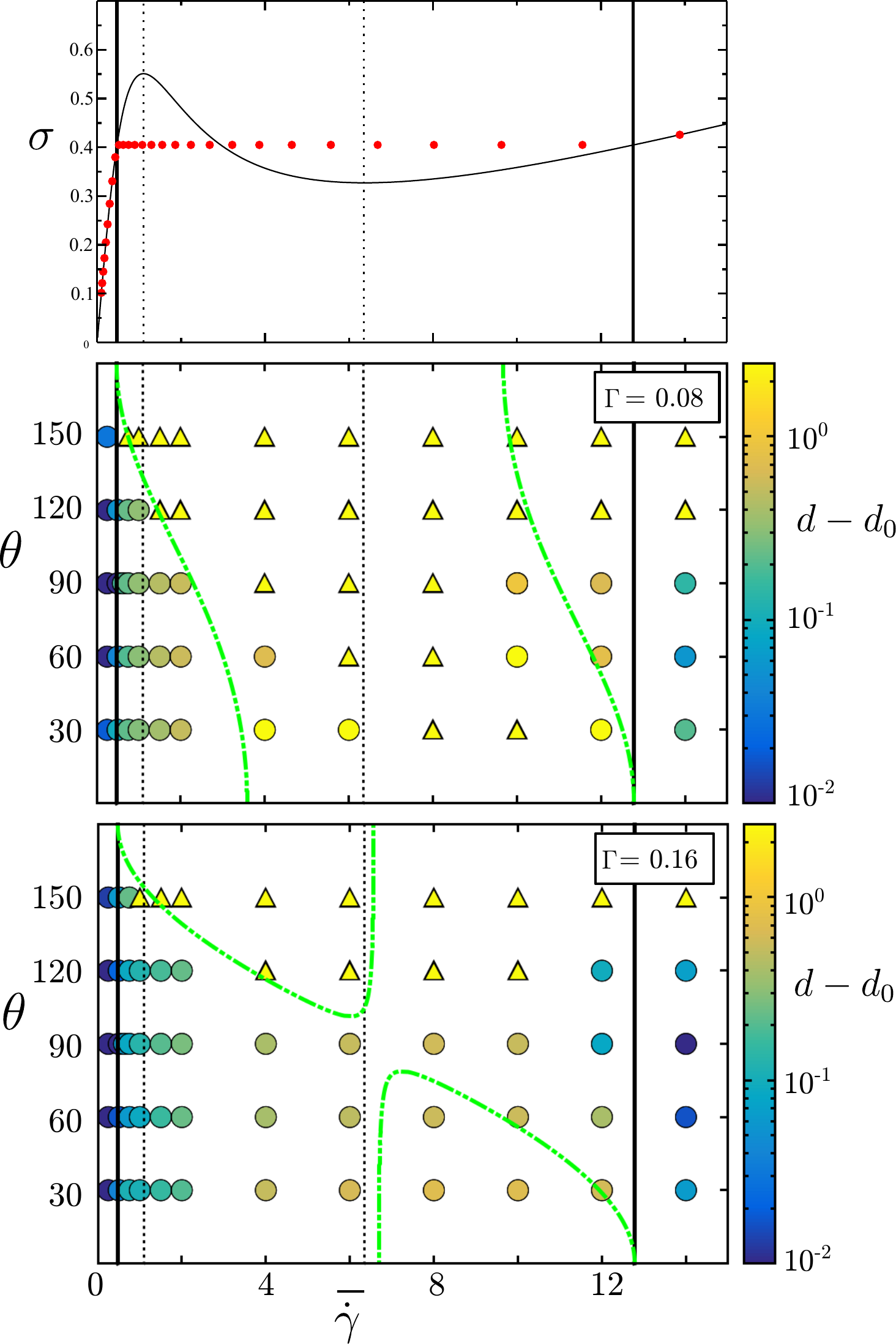}
  \caption{Homogeneous (0D) constitutive curve (black curve) and
  steady state flow curve computed in a 1D simulation (red solid
  circles).  \textbf{Middle:} Colour map of degree of disturbance to
  the fluid-air interface, $d-d_0$, in the plane of the wetting angle,
  $\theta$, and the imposed shear rate, $\gdotb$, for a fixed value of the surface tension $\Gamma=0.08$. \textbf{Bottom:} Corresponding colourmap for $\Gamma=0.16$.
 Green dash-dotted  line shows the analytical prediction of Ref.~\cite{Skorski2011}.
  Johnson-Segalman model, $\etas = 0.025$.}
  \label{fig:PD_wetting}
\end{figure}

The bottom panel of Fig.~\ref{fig:PD_eta0.025} shows results exactly
as in the middle panel, but now for an initial condition that, in the
two windows of shear rate between the solid and dotted vertical lines,
is one of metastable homogeneous shear. At relatively high values of
surface tension in these metastable regimes, the bulk flow remains
unbanded throughout the entire simulation. The threshold for the onset
of a disturbed fluid-air interface is then that which pertains to an
unbanded bulk, as given by the red dashed line, following the criterion
of Hemingway and
Fielding~\cite{Hemingway2017,hemingway2019edge}. (Recall that this
criterion is only valid at low shear rates.) The metastable states of
homogeneous shear accordingly remain unbanded at high enough surface
tension, and with an undisturbed fluid-air interface, leading to black
crosses in Fig.~\ref{fig:PD_eta0.025} (bottom). For a shear banded
initial condition, in contrast, the fluid-air interface must be
disturbed~\cite{Skorski2011}, indeed leading to blue or turquoise
circles in Fig.~\ref{fig:PD_eta0.025} (middle) in place of the black
crosses in the bottom panel.

\subsection{Effect of wetting angle}

Recall that the angle subtended by the fluid-air interface where it
meets the walls of the flow cell defines the wetting angle. The
equilibrium value of this quantity in the absence of any imposed
shear, $\theta$, is set by the boundary conditions for the
Cahn-Hilliard sector of the dynamics (Eqns.~\ref{eq:SI_CH_BCs}).  All
simulations reported so far have taken a value $\theta=90^\circ$, such
that the interface is vertical and flat in the absence of shear. We
consider finally the effect of varying $\theta$. For values of
$\theta\neq 90^\circ$, the fluid-air interface is either concave or
convex even in the absence of shear, with a degree of bowing
$d_0$. Accordingly, we subtract this quantity $d_0$, from the degree
of interfacial bowing in the presence of shear, $d$, to get the degree
of interfacial disturbance caused by shear, $d-d_0$.

Colourmaps of this quantity are shown as a function of wetting angle,
$\theta$, and imposed shear rate, $\gdotbar$, for two different values
of the surface tension of the fluid-air interface, $\Gamma$, in
Fig.~\ref{fig:PD_wetting} (middle and bottom). Values of
$\theta,\gdotbar$ for which the fluid-air interface is found to be
partially edge fractured in our simulations are shown by
circles. Values for which full edge fracture occurs are shown by
triangles. The transition between partially and fully edge fractured
states as predicted analytically by Skorski and
Olmsted~\cite{Skorski2011} is show by the green dashed lines.

The arguments of~\cite{Skorski2011} rested on the shape of the
fluid-air interface determined by force balance, taking into account
the jump in $N_2$ across the interface between the shear bands, the
relative widths of the low and high shear bands (which is set by
$\gdotbar$), and the wetting angle at the walls. As a direct result,
the transition lines predicted in that work show the symmetry apparent
in Fig.~\ref{fig:PD_wetting} in the plane of $\theta$ and
$\gdotbar$. The results of our numerical simulations do not obey even
these basic symmetries, however. This discrepancy again presumably arises due to
the presence of secondary flows near the fluid-air interface, which
lead these basic assumptions of Skorski and Olmsted to break down.

\section{Conclusions}
\label{sec:conclusions}

In this work, we have studied the interplay between shear banding and
edge fracture in complex fluids, mapping out phase diagrams of the
levels of edge disturbance and bulk (or quasi-bulk) shear banding as a
function of the surface tension of the fluid-air interface, the
wetting angle at the walls of the flow cell, and the imposed shear
rate. We have explored in particular the basic result announced in
Ref.~\cite{hemingway2018edge}: that precursors to edge fracture can
induce quasi-bulk shear banding, even in fluids for which the
underlying constitutive curve is monotonic, such that a bulk flow is
predicted to be unbanded in the absence of edge effects. We have also
appraised analytical predictions, made within some assumptions, that
shear banding can induce edge fracture~\cite{Skorski2011}.  Our
simulations have shown that this basic prediction remains valid, but
that the phase boundaries are qualitatively modified due to a
breakdown of those assumptions.

As discussed above, an outstanding question is whether the underlying
constitutive curve of monodisperse linear entangled polymers is
monotonic or non-monotonic, and so whether a steady applied shear flow
should be homogeneous or shear banded. Experimental evidence for shear
banding is mixed and controversial~\cite{Tapadia2006,Hu2007,Hu2010},
with edge fracture often discussed as a confounding
factor~\cite{Sui2007,Li2013,Wang2014,Li2014a,Hu2010}. It would be
interesting to reappraise these existing experiments data in the light
of our new theoretical findings, and to perform new experiments aimed
specifically at studying the interplay between shear banding and edge
fracture described here.

We discuss finally some shortcomings of our work that should be
rectified in future studies.  We have assumed throughout a base state
corresponding to a state of time-independent shear flow (whether
banded or unbanded). Edge fracture is also widely seen in transient
flows such as shear startup. Future theoretical work should therefore
consider the effects of a time-dependent base state on the phenomena
reported here.  We have further considered the limit of planar Couette
flow, arguing that this geometry provides a good approximation to
cylindrical Couette, cone-plate or plate-plate flow. We have thereby
neglected the stress heterogeneity, streamline curvature, and
secondary bulk flows that can arise in such devices. Future simulation
studies should take these effects properly into account. We have also
ignored wall slip, which occurs widely in complex fluids.  Future
theoretical studies should consider the relative dominance of and/or
interplay between edge fracture and wall slip. Finally, we have
ignored inertia, which may be relevant to the non-linear dynamics of
edge fracture, and should be considered in future theoretical work.

{\it Acknowledgements -- } The research leading to these results has
received funding from the European Research Council under the EU's 7th
Framework Programme (FP7/2007-2013) / ERC grant number 279365. 

\clearpage
\bibliography{p2}

\end{document}